\documentclass[preprint,preprintnumbers,amsmath,amssymb,floatfix,nofootinbib]{revtex4}
\usepackage[utf8]{inputenc}
\usepackage{graphicx}
\usepackage{hyperref}
\usepackage{epsf, array, color}
\usepackage{graphicx}
\usepackage{subfigure}
\usepackage{bbold}
\usepackage{slashed}
\usepackage{amsthm}
\usepackage{amsmath}
\usepackage{multirow}

\DeclareMathAlphabet{\mathcalligra}{T1}{calligra}{m}{n}
\newcommand{\beq}{\begin{equation}}
\newcommand{\eeq}{\end{equation}}
\newcommand{\bal}{\begin{align}}
\newcommand{\eal}{\end{align}}
\newcommand{\bit}{\begin{itemize}}
\newcommand{\eit}{\end{itemize}}
\newcommand{\ben}{\begin{enumerate}}
\newcommand{\een}{\end{enumerate}}

\renewcommand{\eqref}[1]{Eq.~(\ref{eq:#1})}

\newcommand{\figref}[1]{Fig.~\ref{fig:#1}}

\newcommand{\tabref}[1]{Tab.~\ref{tab:#1}}

\newcommand{\f}{\frac}

\newcommand{\gev}{{\ \rm GeV}}

\newcommand{\be}{\begin{eqnarray}}
\newcommand{\ee}{\end{eqnarray}}
\newcommand{\bea}{\begin{eqnarray}}
\newcommand{\eea}{\end{eqnarray}}

\newcommand{\thetaW}{\theta_{\rm _W} }

\begin{document}
\pagestyle{plain}
% Title Page
\title{\boldmath Diphoton Excess through Dark Mediators}
\author{Chien-Yi Chen}
\affiliation{Department of Physics and Astronomy, University of Victoria, Victoria, BC V8P 5C2, Canada}
\affiliation{Perimeter Institute for Theoretical Physics, Waterloo, ON N2J 2W9, Canada}

\author{Michel Lefebvre}
\affiliation{Department of Physics and Astronomy, University of Victoria, Victoria, BC V8P 5C2, Canada}

\author{Maxim Pospelov}
\affiliation{Department of Physics and Astronomy, University of Victoria, Victoria, BC V8P 5C2, Canada}
\affiliation{Perimeter Institute for Theoretical Physics, Waterloo, ON N2J 2W9, Canada}

\author{Yi-Ming Zhong}
\affiliation{C.N.~Yang Institute for Theoretical Physics, Stony Brook University, Stony Brook, New York 11794, USA}

\preprint{YITP-SB-16-05}

\vglue 0.5truecm

\begin{abstract}
%In light of recent results of searches for the diphoton resonances by ATLAS and CMS collaborations at 13 TeV where 
%an excesss of around 750 GeV has been reported, we investgate a possiblility that dark mediators 
%mimic real photon signals. Those dark mediators can be dark photons or dark scalars. Numbers of events in various regions of the detector
%due to dark mediators decay are predicted. 

Preliminary ATLAS and CMS results from the first 13 TeV LHC run have encountered an intriguing excess of events in the 
diphoton channel around the invariant mass of 750 GeV. We investigate a possibility that the current excess is due to a heavy resonance 
decaying to light metastable states, which in turn give displaced decays to very highly collimated $e^+e^-$ pairs.
Such decays may pass the photon selection criteria, and successfully mimic the diphoton events, especially at low counts. We investigate two classes of such models, characterized by the following underlying production and decay chains: $gg \to S\to A'A'\to (e^+e^-)(e^+e^-)$ and 
$q\bar q \to Z' \to sa\to (e^+e^-)(e^+e^-)$, where at the first step a heavy scalar, $S$, or vector, $Z'$, resonances
are produced that decay to light metastable vectors, $A'$, or (pseudo-)scalars, $s$ and $a$. Setting the parameters of the 
models to explain the existing excess, and  taking the ATLAS detector geometry into account, we marginalize over the properties of heavy resonances in order to derive the expected lifetimes and couplings of metastable light resonances. We observe that in the case of $A'$, 
the suggested range of masses and mixing angles $\epsilon$ is within reach of several new-generation intensity frontier experiments. 

\end{abstract}

\maketitle
\newpage
\section{Introduction}

The start of the LHC run at 13 TeV center-of-mass energy has brought an unexpected -- from the minimalist point of view -- excess of 
events in the diphoton channel with the invariant mass of about 750 GeV~\cite{ATLAS:2015xxx,CMS:2015dxe}. In the Standard Model (SM) of particles and fields this energy
is not associated with any known resonance, and may be the first sign for elusive New Physics (NP). The appearance of the 
``bump'' in the diphoton spectrum, despite its rather limited statistical significance that may disappear or strengthen 
with more data, has generated a lot of excitement among physicists who wait for {\em any} manifestation of NP beyond SM (BSM) at the weak scale. 

It is true that in most models of NP, the diphoton channel would not necessarily be the ``discovery mode". That is, other manifestations of a 
(tenuous) 750 GeV resonance might have been expected first. Nevertheless, large classes of models where said resonance is produced from the
fusion of the SM gauge bosons and/or quark-antiquark pairs with subsequent decay to the diphoton states have appeared in the literature, 
most of them being tailored for the occasion. 
While the mass of a new resonance suggested by the CMS and (mostly) ATLAS data is to be around 750 GeV, its spin and parity remain 
open for discussion. Spin zero and two resonances come as the most natural candidates, while spin one resonance is disfavored by the 
so-called ``Landau-Yang theorem" that forbids the two photons in any state with the total angular momentum equal to one \cite{Landau:1948kw, Yang:1950rg}. The couplings of the 
spin-zero resonances to photons or gluons cannot be expected to arise at dimension four or lower operator level, and therefore it is reasonable 
to expect that 750 GeV resonance is also coupled to the weak-scale particles, charged under the SM gauge groups. 
The loops of these particles (for example, vector-like fermions~\cite{McDermott:2015sck,Chao:2015nsm,Dev:2015vjd,Murphy:2015kag,Ellis:2015oso,Wang:2015omi,Kawamura:2016idj,Dolan:2016eki,Dutta:2016jqn,Ge:2016xcq}) may have led to the effective couplings of the NP resonance to gauge bosons \cite{Alves:2015jgx,Buttazzo:2015txu,Bai:2015nbs,Ding:2015rxx,Huang:2015evq,Berthier:2015vbb,Gu:2015lxj,Cai:2015hzc,Son:2015vfl,Cheung:2015cug,Badziak:2015zez,Han:2015qqj,Moretti:2015pbj,Huang:2015rkj,
Chakraborty:2015jvs,Han:2015dlp,Ghosh:2015apa,Kanemura:2015vcb,Chiang:2015tqz,Kobakhidze:2015ldh,Altmannshofer:2015xfo,Ahmed:2015uqt,Falkowski:2015swt}. 
If this picture is indeed valid, then more signatures  of weak-scale NP are likely to come from the future data.

While noting still a rather limited significance of the excess, it is reasonable to question every element of the existing anomaly. 
In particular, it is important to ask whether {\em light} BSM final states may be confused with the diphoton signal. A general framework for 
such scenario has been already discussed in several publications  \cite{Agrawal:2015dbf,Dasgupta:2016wxw,Draper:2012xt,Bi:2015lcf}. A heavy resonance $X$ produced by the gluon-gluon or quark-antiquark fusion 
may decay to a pair of light BSM states $Y$ that have weak instability against subsequent decays to electron-positron pairs or photon pairs. 
We will call the $Y$ states as ``dark mediators" (see {\em e.g.} Refs. \cite{Boehm:2002yz,Boehm:2003hm,Pospelov:2007mp,ArkaniHamed:2008qn,Essig:2013lka,Hewett:2012ns}). 
If the decay length of $Y$ is commensurate with the 
linear geometry of the detector ({\em e.g.}, of the inner tracker and eletro-magnetic calorimeter) and its mass is in the MeV-GeV range, then emergent 
highly collimated pairs of photons and/or electron-positron pairs may successfully mimic actual photons. Therefore, the zest of this scenario
is that a new 750 GeV resonance opens the door to the light weakly coupled states coupled to the SM sector, which is a particular realization of the ``hidden valley" idea \cite{Strassler:2006im,Chang:2015sdy,Chang:2015bzc,Bian:2015kjt,Knapen:2015dap,Dey:2015bur,Davoudiasl:2015cuo,Das:2015enc,Kaneta:2015qpf,Jiang:2015oms,Yu:2016lof,An:2015cgp}. 

In recent years,  dedicated searches of light weakly coupled states coupled to electrons, photons, muons and 
other light particles have become an important BSM direction at the intensity frontier \cite{Bjorken:2009mm,Bjorken:1988as,Riordan:1987aw,Bross:1989mp,Batell:2009yf,Strassler:2006im,Strassler:2006qa,Essig:2009nc,Freytsis:2009bh,Essig:2010xa,Blumlein:2011mv,Andreas:2012mt,Pospelov:2008zw,Reece:2009un,Aubert:2009cp,Hook:2010tw,Babusci:2012cr,Archilli:2011zc,Abrahamyan:2011gv,Merkel:2014avp,Dent:2012mx,Davoudiasl:2012ig,Davoudiasl:2012ag,Davoudiasl:2013aya,Endo:2012hp,Balewski:2013oza,Adlarson:2013eza,Agakishiev:2013fwl,Andreas:2013lya,Battaglieri:2014hga,Lees:2014xha,Adare:2014ega,Kazanas:2014mca,Blumlein:2013cua,CERNNA48/2:2015lha}. They have resulted in 
significantly strengthened constraints on the mass--coupling parameter states of light NP particles. 
The purpose of this paper is to explore the consequences of the scenario where a 750 GeV resonance decays to dark mediators 
in terms of its implications for the intensity frontier searches. To that effect, we construct two explicit models, with heavy spin-zero and spin-one resonances, 
that decay to dark mediators. The parameters of the models are chosen to fit the current ATLAS excess of the diphoton events
under the assumption that decaying mediators do indeed pass the selection criteria for the photon identification. In the process, we make 
careful accounting for the ATLAS geometry and the distribution of dark mediators over the effective decay length.
The end result is a suggested range for masses and couplings of dark mediators that falls largely within reach of the next generation of 
intensity frontier experiments ({\em e.g.}~\cite{Abrahamyan:2011gv, Essig:2010xa, Stepanyan:2013tma, Freytsis:2009bh, Balewski:2013oza, Echenard:2014lma, Alekhin:2015byh,Ilten:2015hya}).
%\item Structure of the paper

The paper is organized as follows. We first introduce the theoretical framework for the dark mediator explanation of the 750 GeV 
candidate resonance in \S\ref{sec:TheoreticalMotivation}. We then calculate the strength of 
expected signal, evaluate the probability of light particles decays inside 
the relevant parts of ATLAS detector, and present favored parameter spaces for various models in \S III.
Different experimental strategies that would allow differentiating diphoton from di-dark mediator events are discussed in \S\ref{sec:Discussion}. We  conclude in \S\ref{sec:Conclusion}.

\section{Theoretical motivation}
\label{sec:TheoreticalMotivation}
\subsection{750 GeV Scalar Resonance}
\label{sec:dap}
In this sub-section, we consider a model of a heavy dark scalar (or pseudo-scalar) resonance $S$ produced via gluon fusion that decays to 
the pair of two metastable ``dark photon" particles $A'$. Each $A'$ gives displaced decays to $e^+ e^-$ pairs so that 
the whole chain can be represented as 
\beq gg \to S \to A'A' \to (e^+e^-) (e^+e^-). \eeq
Here we explore a possibility that $m_S \simeq 750$ GeV, but $A'$ is light, $m_{A'} <\mathcal O({\rm few~ GeV})$. 
%$S$ is produced through the gluon fusion channel and then decays into dark photons ($A'$), which then decays into 
%an electron-positron pair. 
Because each dark photon carries a significant fraction of energy of the 750-GeV scalar, the $e^+e^-$ pair from the decay of $A'$
are extremely collimated. The opening angle of $e^+ e^-$ pair is around $2m_{A'}/E_{A'}$, 
where $m_{A'}$ and $E_{A'}$ are the mass and 
energy of $A'$, respectively. For sub-GeV $A'$s this angle is less than 0.01. Therefore it is plausible that events originating from the 
decay of $A'$ could pass the selection criteria for a real photon set by {\em e.g.} The ATLAS collaboration.

Dark photon models have been studied extensively in the literature since the 1980's \cite{Holdom:1985ag, Galison:1983pa}. In recent years, the attention to dark photons have been spearheaded by their possible connection to various particle physics and astrophysics ``anomalies" (see {\em e.g.} \cite{Boehm:2003ha,Pospelov:2007mp,ArkaniHamed:2008qn,Pospelov:2008zw}).  The  minimal dark photon 
model consists of a new massive vector field that couples to the SM $U(1)$ via the so-called kinetic mixing operator,
%We first discuss this in an effective field theory (EFT) approach by introducing an effective Lagrangian as follows,
%
\be
{\cal L}_\text{gauge} &=&-{1 \over 4} B_{\mu\nu} B^{\mu\nu} -{1 \over 4}F'_{\mu\nu} F'^{\mu\nu} + { \epsilon_Y \over 2} F'_{\mu\nu} B^{\mu\nu}+
\frac12 m_{A'}^2 A'_\mu A{'^\mu}
\ee
where $F'_{{\mu\nu}}=\partial_{[\mu} A'_{\nu]}$ and $B_{{\mu\nu}}=\partial_{[\mu} B_{\nu]}$  are field strengths of the $U(1)_D$, $U(1)_Y$ gauge group respectively.
The mass term breaks the $U(1)_D$ explicitly but does not ruin the renormalizability. 
 $\epsilon_Y$ is the kinetic mixing parameter, which we will explicitly assume to be much smaller than one. It dictates the magnitude of the coupling of $A'$ to the SM sector. Even if the boundary conditions in the deep UV are such that $\epsilon_Y(\Lambda_{UV})=0$, the non-zero  mixing can be mediated by a 
loop process with heavy particles charged under both $U(1)$ groups~\cite{Holdom:1985ag}.
In such a scenario, the choice   $\epsilon_Y\ll 1$ is justified due to the expected loop suppression.
%$\epsilon_Y$ is small because it comes from the loop process and thus is veryin a 
After electroweak symmetry breaking (EWSB), the SM gauge field $B_\mu$, and $W^3_\mu$ mix with the new gauge field $A'_\mu$. The resulting mass eigenstate $Z'$ couples to the SM electromagnetic and weak neutral currents. In the limit 
$$m_{Z'} \ll m_{Z},\quad \epsilon_Y\ll 1,$$ 
the mixing between $A'$ and the SM $Z$-boson is negligible, while the coupling between $Z'$ and SM fermions are given by
\beq
\epsilon_Y \cos \theta_W e Q \equiv \epsilon e Q,
\eeq
where we introduce $\epsilon \equiv \epsilon_Y \cos \theta_W$. For more detailed discussions on the kinetic mixing, see Appendix~\ref{km}. Finally, to avoid the proliferation of notations, we will call the physical $Z'$ particle as $A'$, and refer to it
as the dark photon.

\begin{figure}[htbp]
   \centering
   \includegraphics[width=0.4\textwidth]{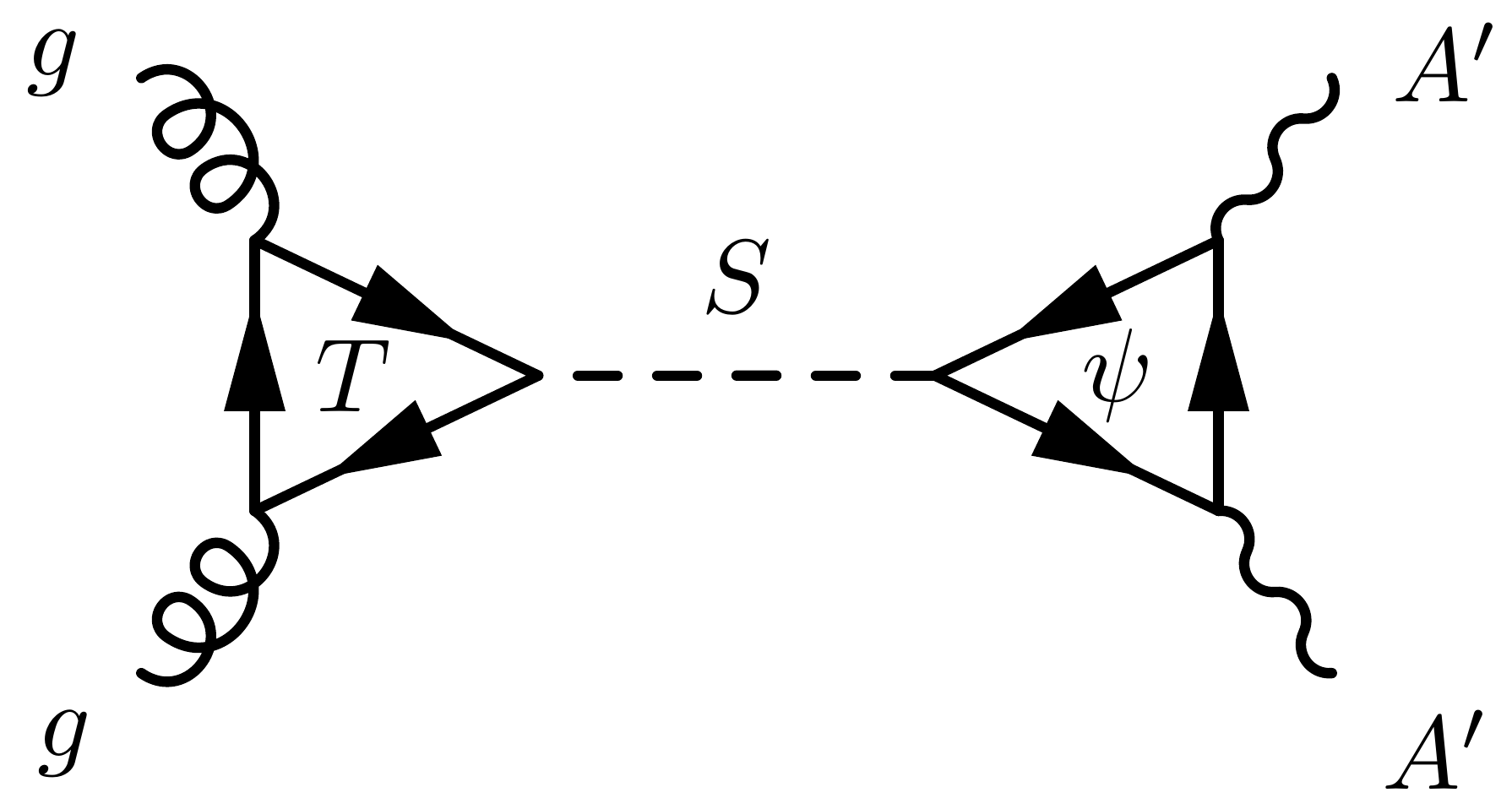} % requires the graphicx package
   \vspace{-5mm}
   \caption{Representative Feynman diagrams for $gg \to S \to A'A' $, where $S$ is the 750 scalar resonance and $A'$ is the light on-shell dark photon that faking photons.}
   \label{fig:diagram1}
\end{figure}

Our goal is to derive the acceptable range for masses and couplings in the proposed scenario. To achieve this, we need to 
specify the couplings of scalar $S$ to gluons and dark photons beyond the effective $\rm{dim} =5$ operators. 
To that effect, we introduce a vector-like colored fermion, $T$,
and a dark fermion, $\psi$, which is a singlet under the SM gauge group.
The resulting Lagrangian reads
\be
{\mathcal L_{S}} = \f{1}{2} (\partial_\mu S)^2 -\f{1}{2}{m^2_S} S^2+\bar{f}i\slashed{D} f+ \bar{T}(i\slashed{D} -m_T)T  - \lambda_T S \bar{T}T + \bar{\psi}(i\slashed{D} -m_\psi)\psi -\lambda_d S \bar{\psi}\psi
\label{L_S}
\ee
where $f$ stands for a generic SM fermion. The covariant derivative here is \beq D_\mu = \partial_\mu -i(g_{d} Q_d + e \epsilon Q_f ) A'_\mu -i e Q_f A_\mu - i g_s G_\mu^a t^a, \eeq 
where
%Here we include the coupling to the  to account interactions between $T$ and gluons. 
$Q_f$ and $Q_d$ are  $U(1)_\text{EM}$ and $U(1)_D$ charges, respectively.
$e$, $g_s$ and $g_d$ are $U(1)_\text{EM}$, $SU(3)_c$, and $U(1)_D$ gauge couplings, respectively. $\lambda_T$ and $\lambda_d$ 
are the Yukawa couplings of $S$ to $T$ and $\psi$ fermions, respectively. 
Notice that one does not have to choose positive parity, and $S \bar{T}i \gamma_5 T$ 
pseudo-scalar couplings could also serve the same purpose.   $T$ and $\psi$ fermion loops mediate the production and decay of 
$S$ resonance, as shown in \figref{diagram1}.

%The origin of the $A'$ is not directly relevant for our work, which can either from a dark Higgs mechanism or Stueckelberg mechanism. 

%

%The process of interest is $$gg \to S \to A'A' \to (e^+e^-) (e^+e^-),$$ where $S$ is produced through the gluon fusion via $T$ and then  decays into a pair of dark photons. Each dark photon then decays into $e^+e^-$ pair, which can mimic the real photon 
%signal. 

Having formulated the model, we are now ready to evaluate the strength of the fake diphoton signal in it. 
We start from the master formula for the signal, 
\be
\sigma_{\rm Signal} = \sigma_{pp\to S} \times {\rm Br}_{S\to A'A'} \times P\left(\left.A'A'\to (e^+e^-)(e^+e^-)\right|\gamma\gamma\right),
\label{eq:mformula}
\ee
where $\sigma_{pp\to S} $ is the cross section for producing 750 GeV resonance $S$ and ${\rm Br}_{S\to A'A'} $ is the branching ratio 
of this resonance decaying to two dark photons. $P\left(\left.A'A'\to (e^+e^-)(e^+e^-)\right|\gamma\gamma\right)$ is the probability that two dark photons 
decay to electron-positron pairs inside the detector (within appreciable distance), passing the selection criteria for the 
diphoton events, and successful reconstruction. It is the most complicated object. It depends on factors such as the detector geometry, the detector acceptance, the reconstruction efficiency, as well as the decay length of $A'$, and the mass of $A'$ that affects the size and the shape of the shower in the EM calorimeter. 
We will abbreviate $P\left(\left.A'A'\to (e^+e^-)(e^+e^-)\right|\gamma\gamma\right)$ as $P_{\text{acc}}$.
The existing excess in the diphoton channel found by ATLAS ~\cite{ATLAS:2015xxx} is at the level of  $\sigma_{\rm Signal}  \simeq 5-10$ fb, which 
corresponds to $\sim$ 16 to 32 events.

\subsubsection*{Production and decay of $S$ in a $U(1)_D$ model} 

Data suggest that the total width of $S$ is around $5-45$ GeV, and therefore the narrow width approximation for $S$ suffices for 
our accuracy. The production cross section of $S$ through gluon fusion is given by
\be
\label{sigma}
\sigma(pp \to S) = {\pi^2 \over 8 s\,m_S }\Gamma(S\to gg) \int_{m_S^2/s}^1 {dx \over x} f_g(x,m^2_S) f_g\left({m_S^2/s \over x},m^2_S\right),
\ee
where $\sqrt{s}=13$ TeV is the center of mass energy and $f_g(x,Q^2)$ is the gluon parton distribution function evaluated at $Q^2$. 
We assume that the decay width of $S \to gg$ entering in (\ref{sigma}) is mediated by the loop of heavy vector-like fermions $T$. 
The actual constraints on $m_T$ would critically depend on $T$-fermion decay channels. To reduce the number of parameters to be scanned, 
we will adopt $m_T = 1~{\rm TeV}$ throughout, which is safe relative to direct searches. Note that for such a massive particle in the loops, 
the form factor of the effective $g-g-S$ vertex does not need to be taken into account. 
A very well known formula for the calculation of the width ({\em e.g.}, see~\cite{Gunion:1989we}) gives
\be
\Gamma(S\to gg)&=& {\alpha_s^2 \over 32 \pi^3}{m_S^3 \over m_T^2}\lambda_T^2 |\tau_T\left[1+(1-\tau_T) f(\tau_T)\right]|^2,
\ee
where $\tau_T= {4 m_T^2/m_S^2}$. 
%Since $S$ does not carry any color charge it couples to the gluons through the vector-like top partner $T$. 
%The current limit on a vector-like top partner is ...~\ref{}. We assume $m_T$ is of order $\sim 1$ TeV. 
In this expression, the invariant function $f(\tau)$ is quite familiar from the Higgs physics literature, 
\be
 f(\tau) = \left\{ \begin{array}{ll}
{\rm arcsin}^2\left( \sqrt{\tau^{-1}} \right), & \hspace{0.5cm} \tau > 1 \\
-\frac{1}{4} \left[ \log\left( \frac{1+\sqrt{1-\tau}}{1-\sqrt{1-\tau}} \right) -i\pi \right]^2, & \hspace{0.5cm} \tau\leq 1
\end{array}\right.\,.
\ee
The cross section (\ref{sigma}) can be further improved by taking into account NLO corrections.
With these expressions, we find that a fiducial value for the 
$\sigma_{pp\to S} $ cross section at $m_T \sim 1$ TeV and $\lambda_T \sim 1$  to be around 40 fb.

The branching ratio of $S$ to dark photons  directly follows from the three decay channels 
of the heavy scalar: $S\to gg, S\to \bar{\psi}\psi$, and $S\to A'A'$,
\be
{\rm Br}_{S\to A'A} = \frac{\Gamma_{S\to A'A'}}{\Gamma_{S\to A'A'} + \Gamma_{S\to gg} + \Gamma_{S\to \bar{\psi}\psi}}.
\ee

If kinematically accessible, the decay of $S$ to dark fermions $\psi$ could be the largest:
\be
\Gamma(S \to \bar{\psi} \psi) = \frac{\lambda_d^2 }{8 \pi}m_S \left(1-\frac{4m_\psi^2}{m_S^2}\right)^{3/2}.
\ee
The $S\to A'A'$ decay is induced by the $\psi$ loop and is given by
\be
\Gamma(S\to A'A')&=& {\alpha_d^2 \over 64 \pi^3}{m_S^3 \over m_\psi^2}\lambda_d^2 \left|\tau_\psi\left[1+(1-\tau_\psi) f(\tau_\psi)\right]\right|^2, 
\ee
where $\tau_\psi = {4 m_\psi^2/m_S^2}$. Note that in this expression we have taken $m_{A'} $ to zero, as it is negligibly small 
compared to $m_S$ and $m_\psi$.

\begin{figure}[t]
     \includegraphics[width=0.48\textwidth]{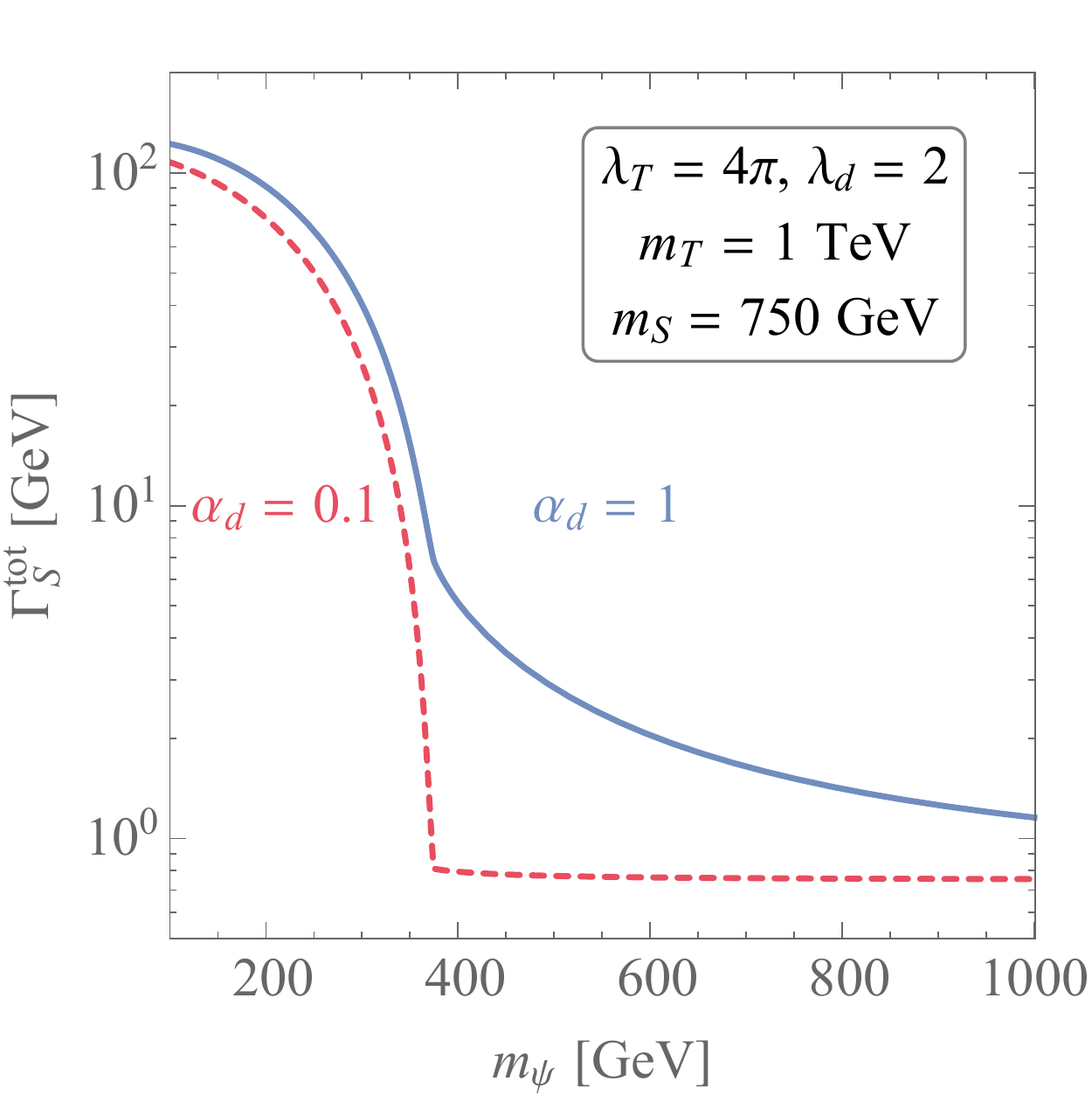}~\includegraphics[width=0.48\textwidth]{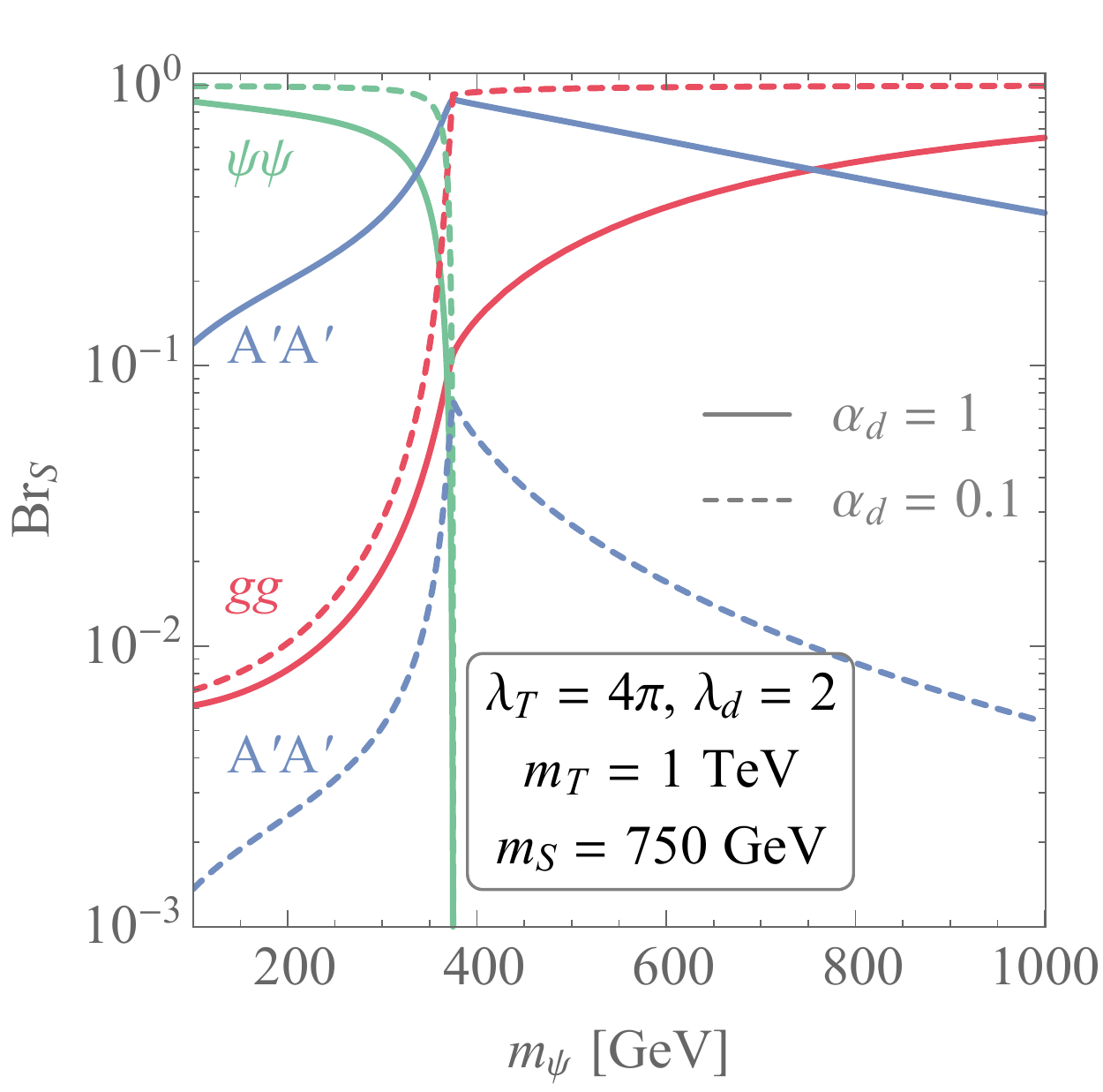} % requires the graphicx package
  \caption{{\em Left panel}: Total decay width of $S$ with $\alpha_d=1$ (blue, solid) and 0.1 (red, dashed). The values of other parameters used here are 
$m_T = 1~{\rm TeV}$, $\lambda_d = 2$ and $\lambda_T $ is taken to the strong interaction limit,  $\lambda_T = 4 \pi$. {\em Right panels}: Branching ratios of $S$ for $S\to gg$ (red), $S\to \bar{\psi}\psi$ (green), and $S\to A'A'$ (blue) with $\alpha_d=1$ (solid) and 0.1 (dashed), with the same choice of other parameters as in the left panel.}
  \label{fig:BRdap}
\end{figure}

The total width and branching ratios of the $S$-resonance are illustrated in Fig. \ref{fig:BRdap}. We have taken $\lambda_T$ to what we will 
consider its uppermost value, $4\pi$, (which would imply a strongly interacting $S-T$ sector). We can observe that if the decays to dark fermions $\psi$ are allowed, one could easily achieve 
a width of the $S$ resonance of $\sim 40~{\rm GeV}$. Rather large branching ratios to pairs of dark photons can be achieved for $\alpha_d \sim \mathcal O(1)$. 
We note in passing that the hierarchy of mass scales, $m_\psi \gg m_{A'}$ and large coupling constant $\lambda_d$  
will create a variety of interesting effects for the dark matter phenomenology, should $\psi$ remain stable on cosmological time scales
(see {\em e.g.} \cite{ArkaniHamed:2008qn,Pospelov:2008jd}).

We now come to the most technically challenging part, the evaluation of $A'$ decays mimicking the 
diphoton signal, $P_{\text{acc}}$. In a hypothetical limit of an infinite detector 
with 100\% efficiency and 100\% faking rate for a dark photon as a regular photon, this probability is simply 
$({\rm Br}_{A'\to e^+e^-})^2$. The branching of dark photons to electrons is well-known \cite{Batell:2009yf}, and is 100\% below the dimuon threshold,
while the $A'$ width is given by 
\be
\Gamma_{A'\to l^+l^-}= \frac{\epsilon^2 \alpha}{3} m_{A'}\sqrt{1-\frac{4m_l^2}{m_{A'}^2}}\left(1+\frac{2m^2_l}{m_{A'}^2}\right).
\ee
At higher $m_{A'}$ one has to include muon and hadronic decay channels, {\em i.e.}, $\Gamma_{A'} = \Gamma_{A'\to e^+e^-} + \Gamma_{A'\to \mu^+\mu^-} 
+ \Gamma_{A'\to \text{hadronic}}$. 

In practice, of course, there are strict geometric requirements where the decays of the dark photons must occur 
so that they can be confused with a real photon. 
Obviously, a very important requirement is that both dark photons decay 
before or inside the first layer of the EM calorimeter, which depends rather sensitively 
on the decay length. 
Suppose that the parent  $S$ particle is produced almost at rest, and then decays into two dark photons,
each of them carries energy around $m_S/2$, where $m_S$ is the mass of the heavy scalar $S$. 
Then the decay length of the dark photon can be written as
\be
L_{A'}(\epsilon,m_{A^\prime})  &=& \gamma_{A'} \beta_{A'} \tau_{A^\prime}(\epsilon,m_{A^\prime}) =  {m_S \over 2 m_{A^\prime}}\sqrt{1-{4 m_{A'}^2 \over  m^2_{S}}} \Gamma^{-1}_{A^\prime},
\ee
 where $\beta_{A'}$ is the velocity of $A'$ observed in the fixed laboratory frame and $\gamma_{A'}\equiv 1/\sqrt{1-\beta_{A'}^2}$ is the boost factor of $A'$. 
$\tau_{A^\prime}(\epsilon,m_{A^\prime}) $ is the lifetime of $A'$ in its rest frame. 
Evidently, $\gamma_{A'} \gg 1$ and $\beta_{A'}$ is almost one. The decay length follows an approximate scaling 
\beq L_{A'}\propto (\epsilon m_{A'})^{-2 }\times m_S \eeq
with largest deviations of this scaling at $m_{A'} \sim m_\rho$. 
Below the dimuon threshold, we have the following useful expression,
\be
L_{A'}(\epsilon,m_{A^\prime}) = 30 ~{\rm cm} \times \left( \frac{m_S}{750~{\rm GeV}}\right) \times \left( \frac{100~{\rm MeV}}{m_{A'}}\right)^2 \times 
\left( \frac{10^{-4}}{\epsilon} \right)^2 .
\label{dap}
\ee

These numbers immediately tell us that {\em currently allowed} region of the dark photon parameter space can 
indeed be compatible with dark photons decaying within reasonable distance inside the LHC detectors
so that they can be confused with real photons. If initial boost distributions of $S$ particles, and angular dependences of  its production 
and of detector geometry could have been neglected,  then $P_{\text{acc}}$ would be determined 
by the relation between some relevant length scale of the detector,  $L_{\rm det}$,  and $L_{A'}$. 
\be
 P_{l<L_{\rm det}}  \propto 1 - \exp\{- L_{\rm det}/L_{A'} \}~~ \Longrightarrow ~~ P_{\text{acc}} \propto ({\rm Br}_{A'\to e^+e^-})^2\times (P_{l<L_{\rm det}} )^2,
\label{acc_naive}
\ee
where $P_{l<L_{\rm det}}$ is the probability of a single photon to decay inside $L_{\rm det}$. This is of course a very crude 
formula that has to be carefully augmented for the detector geometry, boosts, and other factors, 
which we will attempt to do in section \S\ref{sec:prob}. We also note that should one of the dark photons decay 
outside the detector, this would mimic the mono-photon signal with the probability that scales 
as \beq P_{\rm mono} \propto 2\times {\rm Br}_{A'\to e^+e^-}\times P_{l<L_{\rm det}}\times P_{l>L_{\rm det}},\eeq
setting up the stage for an important constraint that would come from corresponding searches. 

%\begin{figure}[t]
 %     \includegraphics[width=0.5\textwidth]{totwidS} % requires the graphicx package
  % \caption{Total width of $S$ with $\alpha_d=1$ (black, solid) and 0.1 (red, dashed). The values of other parameters used here are $m_T = 100, \lambda_T = 4 \pi$, and $\lambda_d = 2$.}
 %  \label{fig:totwidS}
%\end{figure}
%
%\begin{figure}[t]
 %  \includegraphics[width=0.45\textwidth]{BRdap1} % requires the graphicx package
  %    \includegraphics[width=0.45\textwidth]{BRdapp1} % requires the graphicx package
  % \caption{Branching ratios of $S$ for $S\to gg$ (red, dotted), $S\to \bar{\psi}\psi$ (black, dashed), and $S\to A'A'$ (blue, solid) with $\alpha_d=1$ (left) and 0.1 (right). The values of other parameters used here are $m_T = 100, \lambda_T = 4 \pi$, and $\lambda_d = 2$.}
  % \label{fig:BRdap}
%\end{figure}

%\nonumber\\
%f(\tau)&=& (\sin^{-1}(\sqrt{1/\tau}))^2
%f(\tau) &=& \left\{ \begin{array}{ll}
%{\rm arcsin}^2\left( \sqrt{\tau^{-1}} \right), & \hspace{0.5cm} \tau > 1 \\
%-\frac{1}{4} \left[ \log\left( \frac{1+\sqrt{1-\tau}}{1-\sqrt{1-\tau}} \right) -i\pi \right]^2, & \hspace{0.5cm} \tau\leq 1
%\end{array}\right.\,
%\ee
%where $\tau_\psi= {4 m_\psi^2/m_S^2}$. 
%The decay width of $S \to \bar{\psi} \psi$ is as follows,

\subsubsection*{Variations on the dark photon model}
In this subsection we would like to note that the dark photon model is not the only
possibility for a weakly unstable light vector particles. Indeed, there are other UV complete choices 
based on anomaly-free symmetries, such as $B-L$, $L_{L1}-L_{L2}$ (where $L1$ and $L2$ stand for different 
lepton flavors) etc. 
 If we take, for example, a model with $U(1)$-gauged $L_e - L_\tau$ symmetry, then the main couplings of its gauge boson $V$ to 
leptons are 
\be
\label{Vcoupling}
{\cal L} = g_{L_e-L_\tau} V_\alpha \Big( \bar\nu_e \gamma^\alpha \nu_e -\bar\nu_\tau \gamma^\alpha \nu_\tau  + \bar e\gamma^\alpha e- \bar{\tau} \gamma^\alpha \tau\Big)~.
\ee
Here, $g_{L_e-L_\tau}$ is the $U(1)_{L_e - L_\tau}$ gauge coupling, so that the coupling to electrons is rescaled compared to the 
dark photon case as $e\epsilon \to g_{L_e-L_\tau}$. In the entire mass range from a few MeV to $3.6$~GeV the vector boson 
$V$ decays to electrons and neutrinos, with equal probabilities so that ${\rm Br}_{V\to e^+e^-} = 0.5$. 

Despite the fact that one can choose $g_{L_e-L_\tau}$ in the same range as $e\epsilon$ and thus adjust the decay length of $V$ to be commensurate with 
$L_{\rm det}$, this model does not look as a good candidate to mimic the diphoton signal, for the following reasons. 
Firstly, $g_{L_e-L_\tau}$ is required to be very small, $g_{L_e-L_\tau} < 10^{-2}$, from the decay length requirements, which would correspond to a 
tiny $\alpha_d$. This in turn would require some additional model-building to generate an appreciable branching of $S$ 
to $VV$ states. Another reason is that this model will give a non-removable mono-photon signal due to the decay to neutrinos 
 at a rate more than twice the diphoton signal, for any ratio of $L_{\rm det}/L_V$. 
%~\cite{ATLAS:2014hga,Khachatryan:2014rwa}.
On account of these two difficulties, we will abandon further investigation of $U(1)_{L_e - L_\tau}$ models in connection to the 750~GeV resonance.

\subsection{750 GeV Vector Resonance}
If light unstable particles can indeed fake real photons at the LHC,  new possibilities for the spin of the 750 GeV resonance open up. 
In this section we will consider an option of dark mediators being scalar and pseudo-scalar, while 
the decayed 750 GeV resonance being a spin-1 vector boson. Notice that this is a novel possibility 
bypassing the Landau-Yang theorem (see {\em e.g.} earlier related discussion in Ref. \cite{Toro:2012sv,Chala:2015cev}). 

\begin{figure}[htbp]
   \centering
   \includegraphics[width=0.4\textwidth]{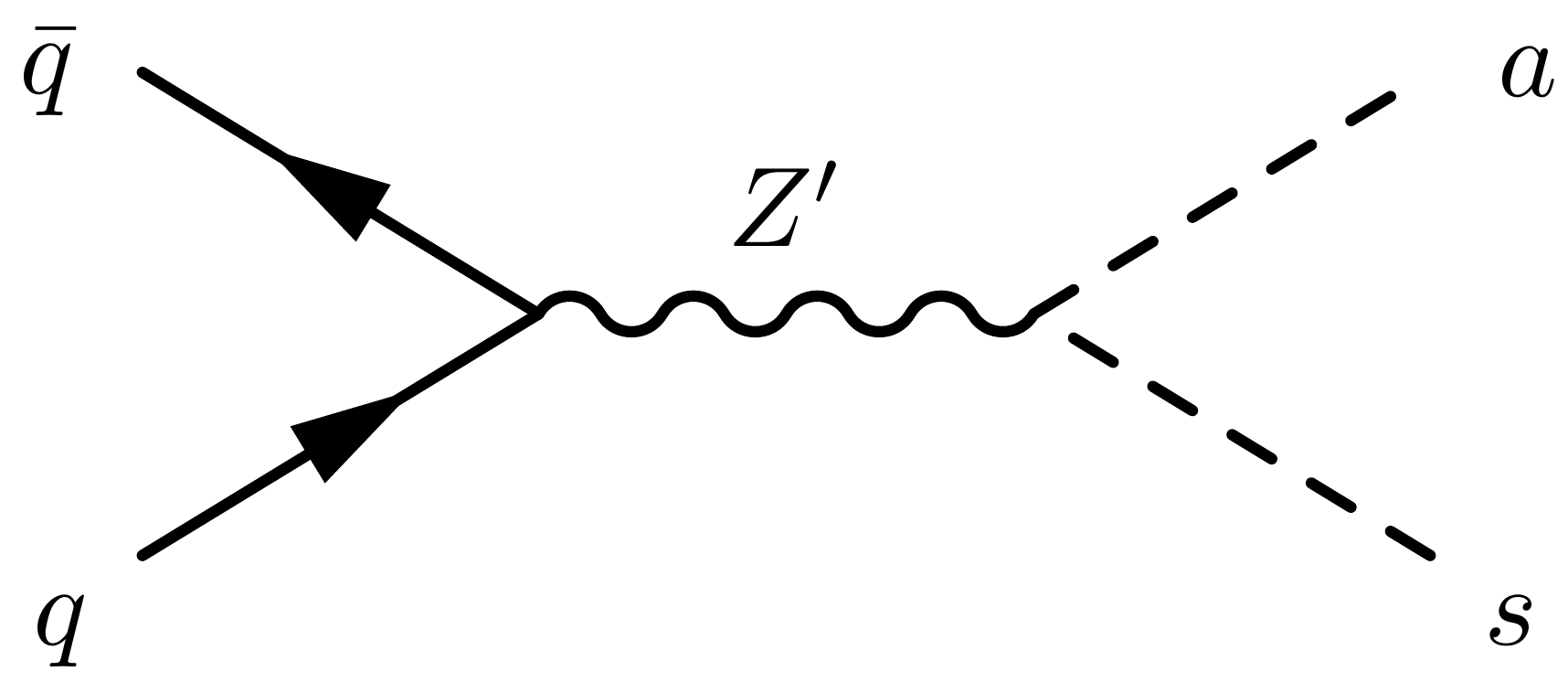} % requires the graphicx package
   \vspace{-4mm}
   \caption{Feynman diagram for $q \bar q \to Z' \to s a $, where $Z'$ is the 750 GeV vector resonance and $s$($a$) is the light on-shell scalar (pseudo-scalar) that faking photons.
%\to (e^+ e^-)(e^+ e^-)$. 
}
   \label{fig:diagram2}
\end{figure}

The scenario of this section is based on the following sequence,
\be
\label{Z'scenario}
q \bar q \to Z' \to s a \to (e^+ e^-)(e^+ e^-),
\ee
where all new particles $Z',s,a$ are assumed to be singlets under the SM gauge group. 
Scalar $s$ and pseudo-scalar $a$ can be combined in a  complex scalar field
\beq \mathcal S = s+ia,\eeq 
that we assume is charged under some new $U(1)_D$ group  with dark charge $Q_d=1$. 
The mass of  a heavy dark $Z'$ boson is taken around 750~GeV. The coupling of  $Z'$ 
to the SM can again proceed via the kinetic mixing operator. To avoid confusion with the 
case of the previous section, we will call the heavy boson $Z'$ (while the light one is $A'$). The Feynman diagrams for the process is shown as \figref{diagram2}.

%is secluded from the SM sector, 
The kinetic mixing operator will couple the $Z'$ to hypercharge of the SM particles (as opposed the electric charge in case of small 
vector mass). Since for the chosen $m_{Z'}$ mass scale
 $$m_{Z'}\gg m_Z,\quad \epsilon_Y\ll 1,$$ 
the coupling between $Z'$ and SM fermions are given by
\beq
\epsilon_Y g' Y = \f{\epsilon e Y}{\cos^2 \theta_W}
\eeq
See Appendix~\ref{km} for more details. 
%We take this limit for the 750 GeV $Z'$ in rest of the section.

The resulting effective Lagrangian reads
\be
{\mathcal L_{Z',\text{eff}}}\supset  \frac{1}{2} {M_{Z'}}^2 {Z'_\mu}^2 + {\bar f_{L,R}} i\slashed{D}f_{L,R} + \left|D_{\mu}\mathcal S\right|^2- m_{\mathcal S}^2  \left|\mathcal S \right|^2 +{\cal L}_{\rm dec},
\ee
where \beq D_\mu = \partial_\mu -i e Q_{f} A_\mu - i\frac{g}{\cos \theta_W} \left(T^3_{L,R} \cos^2 \theta_W- Y_{L,R} \sin^2 \theta_W\right) Z_\mu  -i \left(g_{Z'} Q_d +  \f{\epsilon e Y_{L,R}}{\cos^2 \theta_W} \right) Z'_\mu
\eeq and $f_{L,R}$ includes all left-handed/right-handed SM quarks and leptons.
$Q_{f}$, $T^3_{L,R}$ and $Y_{L,R}$ represent their $U(1)_\text{EM}$, $SU(2)_L$, and $U(1)_Y$ charges respectively. $e$, $g$ and $g_{Z'}$ are electric coupling, weak coupling and the $U(1)_D$ gauge coupling, respectively. ${\cal L}_{\rm dec}$  is the most ``delicate" part of the Lagrangian 
that is responsible for the decays of $a$ and $s$ particles. 
Notice that one cannot simply write down $\lambda_s s \bar{e} e $ and 
$ \lambda_a a \bar e i \gamma_5 e$ operators at the fundamental level, as they would explicitly 
violate both the SM and $U(1)_D$ gauge invariances. Nevertheless these operators can be in fact generalized to 
the following gauge invariant structures of higher dimension: 
\be
\label{effYuk}
{\cal L}_{\rm dec} = \lambda_s s (\bar{e} e) + \lambda_a a (\bar e i \gamma_5 e) ~~\to~~ \frac{{\cal S} \Phi}{\Lambda_\Phi^2} \times  (\bar{L}_1 E_1 H)  +(h.c.).
\ee
In this formula, $L_1$ and $E_1$ are the first generation left- and right-handed lepton fields, $H$ is the SM Higgs field bi-doublet, and $\Phi$ is the 
Higgs field of the dark sector with the charge of $-1$. For the purpose of our discussion, $H$ and $\Phi$ can be replaced by their vacuum expectation values, $v/\sqrt{2}$ and $v_d/\sqrt{2}$. $\Lambda_{\Phi}$ is some energy scale normalizing this effective operator, that now defines the effective Yukawa couplings as 
\be
\lambda_{\cal S} \equiv \lambda_s=\lambda_a = \frac{vv_d}{2\Lambda_\Phi^2}.
\ee
Since it is clear that displaced decays are only possible for $\lambda_{\cal S} \ll 1$ and typically as small as $10^{-4}$ 
while heavy $m_{Z'}$ implies a large dark vev $v_d$,
the scale $\Lambda_\Phi$ can be well above the LHC energy reach. We leave it at that, without trying to provide further 
UV completion to the effective operator (\ref{effYuk}). A further uncertainty in this approach arises from 
a possibility of nontrivial lepton flavor structure of (\ref{effYuk}). To avoid possible complications, we will assume that these couplings are
flavor-diagonal, and will limit $m_{s,a}$ to be below the dimuon threshold.  

%The process of interest is $$ q \bar q \to Z' \to s a \to (e^+ e^-)(e^+ e^-),$$ where  $s$ and $a$ are light mediators with $m_{s,a} \lesssim 2 m_\mu$.

\subsubsection*{Production and decay of $Z'$}

Going over to the production mechanism, we notice, of course, that  $Z'$ does not couple to gluons, and have to be produced in  $q\bar{q}$ fusion. Although the probability of finding (anti-)quarks inside the proton at high energy is smaller compared to
that of gluons, the leading order contribution of this process is at tree-level 
and thus the cross section can be comparable to gluon-initiated but loop-suppressed
processes. 
The production cross section of $Z'$ reads
\be
\label{sigmazp}
\sigma(pp \to Z') =  \int_0^1 dx_1 \int_0^1 dx_2  f_q(x_1,m_{Z'}^2) f_{\bar{q}} (x_2,m_{Z'}^2)
\hat{\sigma}\left[q\left(x_1 {\sqrt{s}\over 2}\right) \bar{q}\left(x_2 {\sqrt{s}\over 2}\right) \to Z'\right],
\ee
where $\sqrt{s}=13$ TeV is the center of mass energy and $f_q(x,Q^2)$($f_{\bar q}(x,Q^2)$) is the quark (anti-quark) parton distribution function evaluated at $Q^2$. 
At the same time, the increase in parton luminosity between run I and run II for the production of the 750 GeV resonance is 
less pronounced for $q \bar q$ compared to gluons, by about a factor of order 3. 
The decay channels of $Z'$ are similar to those of the SM $Z$-boson but with an additional channel, $Z' \to sa$ available in this model.
The decay width to the SM fermions is given by
\be
\Gamma(Z'\to f\bar{f})= {N_c \over 12 \pi } m_{Z'}\left({\epsilon e \over \cos^2\thetaW}\right)^2 \sqrt{1- {4 m_f^2 \over m_{Z'}^2}}\left[ \f{Y_L^2 + Y_R^2}{2}
+{m_f^2 \over m_{Z'}^2}\f{6Y_L Y_R - Y_L^2 -Y_R^2}{2} \right], 
\ee
where $N_c$ represents the number of colors of the SM fermions ($f$) and $Y_L(Y_R)$ stands for the hypercharge of the left-handed (right-handed) SM fermions. 
The decay width of the ``dark" $sa$ channel is
\begin{equation}
\Gamma(Z'\to sa)= {g_{Z'}^2 \over 48 \pi} m_{Z'} \left(1-{2(m_s^2 +m_a^2)\over m_{Z'}^2}+ {(m_s^2 -m_a^2)^2\over m_{Z'}^4} \right)^{3/2}\simeq\f{g_{Z'}^2}{48\pi} m_{Z'}^2.
\end{equation}
We take the limit $m_{s,a}\ll m_{Z'}$ in the second equality. 

\begin{figure}[t]
     \includegraphics[width=0.48\textwidth]{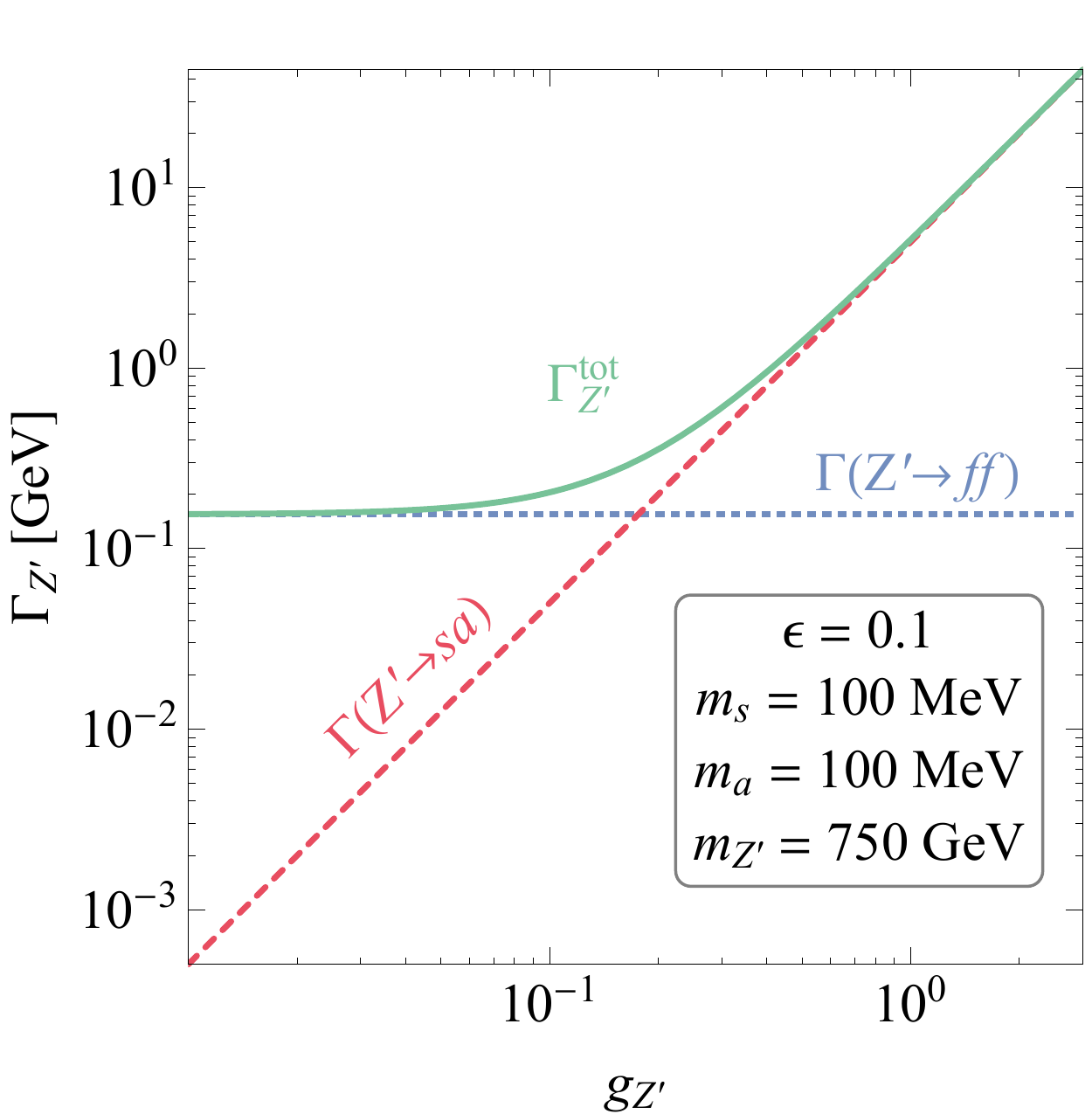}\\% requires the graphicx package
  \caption{Total decay width of $Z'$ (green, solid) along with its partial widths $\Gamma(Z'\to f\bar{f})$ (blue, dotted) and $\Gamma(Z'\to s a)$ (red, dashed) for $\epsilon=0.1$ and 
$m_s=m_a= 100$ MeV. $\Gamma(Z'\to f\bar{f})$ is independent of $g_{Z'}$ since it is only a function of $\epsilon$ whereas $\Gamma(Z'\to s a)$ is proportional to $g_{Z'}^2$.}
  \label{fig:widthZp}
\end{figure}

Fig.~\ref{fig:widthZp} shows the total width of $Z'$ (green, solid) as well as its partial widths $\Gamma(Z'\to f\bar{f})$ (blue, dotted) and $\Gamma(Z'\to s a)$ (red, dashed) for $\epsilon=0.1$ and $m_s=m_a=100$ MeV. 
 %$f$ stands for all the fermions in the SM. 
 $\Gamma(Z'\to f\bar{f})$ does not vary with $g_{Z'}$ since it only depends on $\epsilon$ while $\Gamma(Z'\to s a)$ is proportional to 
$g_{Z'}^2$ and therefore grows with $g_{Z'}$. One can also see that for small $g_{Z'}\sim 0.01$ the dominant decay branching ratio is from $Z'\to f\bar{f}$ and the total width of $Z'$ is also very small.
However, for a large enough $g_{Z'}\sim 3$, not only the dominant channel becomes $Z'\to s a$, but also the width of $Z'$ can reach $\sim 45$~GeV due to $Z' \to sa$ decays without any difficulty. Therefore in the following analysis we use  $g_{Z'}\sim 3$ as a representative point. Also notice that since the branching ratio of $Z'\to s a$ is close to 1 at that point, the parameter $g_{Z'}$ cancels
in the branching ratio and has very small effect on subsequent considerations. 
 
The decay lengths of $s$ and $a$ are as follows,
\be
L_s(\lambda_s,m_s)={m_{Z'}\over 2 m_s}\sqrt{1-{4m_s^2\over m_{Z'}^2}} \Gamma_s^{-1}, \\
L_a(\lambda_a,m_a)={m_{Z'}\over 2 m_a}\sqrt{1-{4m_a^2\over m_{Z'}^2}} \Gamma_a^{-1},
\ee
with 
\be
\Gamma_s = {m_s \over 8 \pi} \lambda_s^2\left(1-{4 m_e^2 \over m_s^2}\right)^{3/2}, \\
\Gamma_a = {m_a \over 8 \pi} \lambda_a^2\left(1-{4 m_e^2 \over m_a^2}\right)^{1/2},
\ee
where $\Gamma_s$ and $\Gamma_a$ are total widths of $s$ and $a$, respectively. We only explore the region below the dimuon threshold
so that one can have Br$_{s\to e^+e^-}$ and Br$_{a\to e^+e^-}$ of order one. 
%Similar to the discussion in \S~\ref{sec:dap} we have the following useful expression,
%\be
%L_{s/a}(\epsilon,m_{s/a}) = 2 ~{\rm cm} \times \left( \frac{m_{Z'}}{750~{\rm GeV}}\right) \times \left( \frac{100~{\rm MeV}}{m_{s/a}}\right)^2 \times 
%\left( \frac{10^{-4}}{\lambda_{s/a}} \right)^2.
%\ee
%This 

%Finally, notice that the width of $Z'$ can reach $\sim 45$~GeV due to $Z' \to sa$ decays without any difficulty. 
%The vertices for the $Z'$ coupling to scalar is 

%\section{S is produced by mixing with SM Higgs}
%In this section we present a model where S is produced through mixing with SM Higgs instead of effective coupling 
%$\kappa_G S G^a_{\mu\nu} G^{a,{\mu\nu}}$. Since the coupling 

%\subsection{Dark photon with a real dark scalar and a dark fermion}

\section{Results}
\label{sec:prob}

\subsection{Geometry of LHC relevant for the diphoton signal}
\begin{figure}[t]
   \centering
   \includegraphics[width=0.8\textwidth]{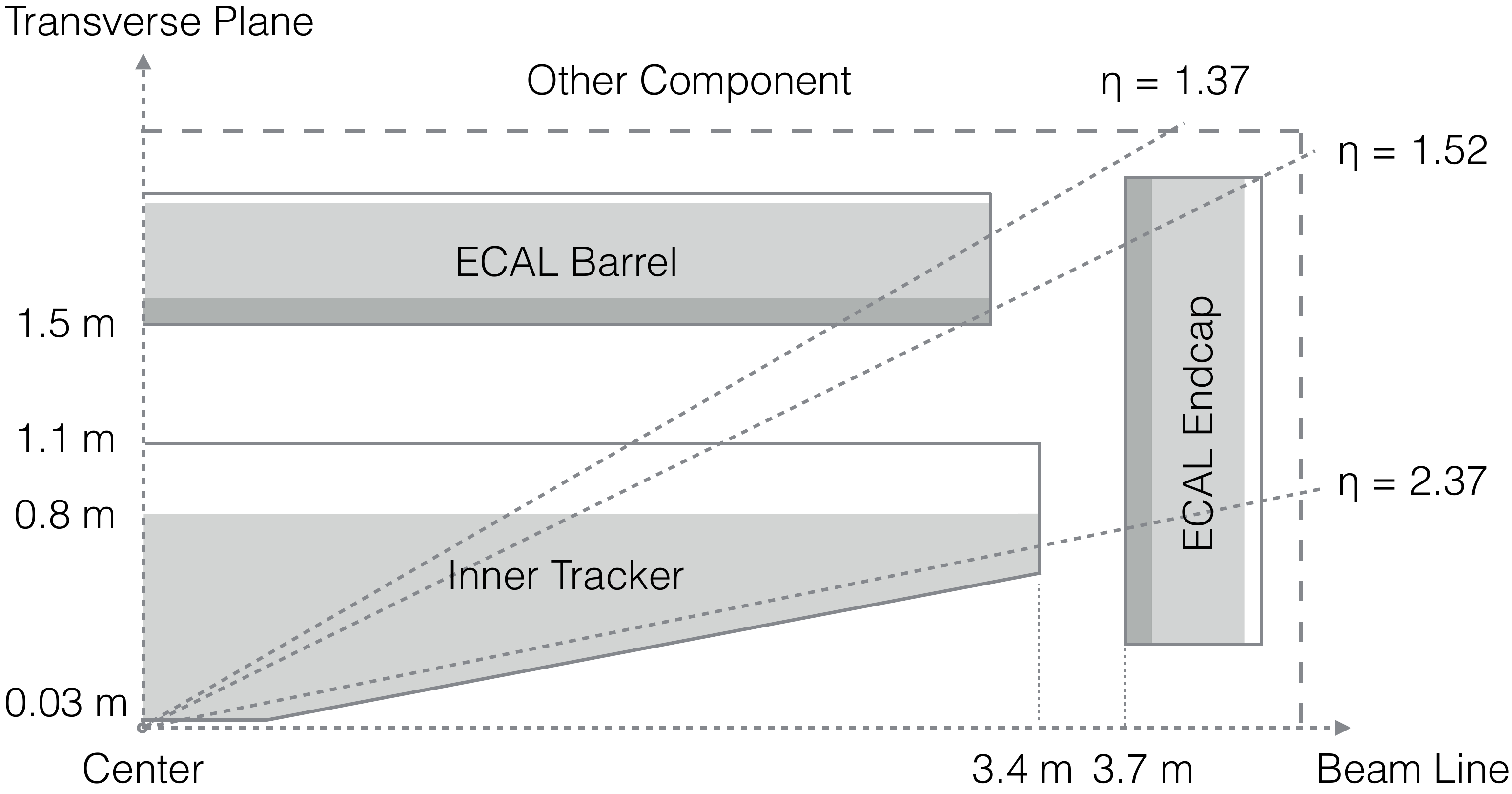} % requires the graphicx package
   \caption{Relevant geometry of the ATLAS detector. Here we only show the  configuration of the inner tracker and ECAL  (1/4 of the $z$ view). Other components of the detector are not shown. The fiducial region of inner detector and ECAL are shaded with gray. The ECAL consists of three layers (see text for details), which we shaded with different tones of gray. We also specific values for the relevant geometry that used in the analysis. Other relevant parameters of the detector can be found in~\tabref{geometry}.}
   \label{fig:geometry}
\end{figure}

The ATLAS detector can be viewed as a series of ever-larger concentric cylinders around the beam line. From the inner region to the outer region, the main detector elements are a silicon pixel and strip tracker, an electromagnetic calorimeter (ECAL), a hadron calorimeter (HCAL), and a muon spectrometer. A $1/4$ of the $z$ view of the detector is demonstrated in \figref{geometry}.

The inner detector tracking system is used to reconstruct primary 
vertices up to a radius in the transverse plane ($r$) less than 0.8 m \cite{ATLAS:2012ana}. Recently, ATLAS has upgraded
the inner detector system and inserted another layer, the Insertable B-Layer (IBL) \cite{Rohne201318}, near the beam-pipe with $0.03$ m $< r <0.04$ m to enhance the tracking
ability and overcome the increased pileup at the LHC run-II. Therefore we define the fiducial volume of the inner detector to be 
in the region $0.03$ m $< r< 0.8$ m. A photon passing through the fiducial volume of the inner detector can convert into an electron-positron pair, which leaves tracks in the fiducial volume. As a result, such photons are classified as converted photon candidates by the ATLAS collaboration.

% Requires the booktabs if the memoir class is not being used
\begin{table}[htbp]
   \centering
   %\topcaption{Table captions are better up top} % requires the topcapt package
   \begin{tabular}{@{} lc @{}} % Column formatting, @{} suppresses leading/trailing space
      \hline
      \multicolumn{2}{c}{ATLAS detector} \\
      \hline
            \multirow{3}{*}{Inner tracker}    & Region: $0.03 ~\text{m} < r < 1.1~\text{m}$\\
            & Fiducial:  $0.03 ~\text{m} < r < 0.8~\text{m}$\\
                  & $|\eta|< 2.5$  \\
                        \hline
      \multirow{3}{*}{ECAL Barrel (EB)}    & Region: $1.15~\text{m} < r <2.25~\text{m}$   \\
      & Fiducial: $1.5~\text{m} < r <1.93~\text{m}$   \\
                  & $|\eta|<1.48$ \\
      \hline
      \multirow{2}{*}{ECAL Endcap (EE)}          &Region: $3.4~\text{m} < |z| < 6.57~\text{m}$  \\
       & Fiducial: $3.7~\text{m} < r <~4.13\text{m}$   \\
                      & $1.38<|\eta|<3.2$   \\
                      \hline
   \end{tabular}
   \caption{Geometric parameters and fiducial regions of inner trackers and ECALs of ATLAS. $r$ denotes the transverse radius from the beam line. $z$ denotes the distance from the center of the detector along the beam line. $\eta$ denotes the pseudorapidity with respect to the center of the detector. }
   \label{tab:geometry}
\end{table}

The ECAL  (as well as HCAL) is composed of a barrel and two endcaps. The ATLAS ECAL is a lead-liqid argon sampling calorimeter. The relevant geometrical parameters of the ECAL components are summarized in \figref{geometry} and \tabref{geometry}~\cite{Aad:2009wy}.
The ECAL consists of three layers, starting at $r = 1.5$ m.  A photon is categorized as an unconverted photon candidate if it converts inside the region between 0.8 m and 1.5 m, consisting of the final part of the tracking system and a gap between the inner tracker and the first layer of the ECAL, since it does not leave any reconstructible tracks. In summary, the fiducial volumes of the event reconstruction for the converted and unconverted photons are $0.03$ m$ < r< 0.8$ m and $0.8$ m$ < r<1.6$ m, respectively. Note that if the second layer of the ECAL is also included, the fiducial volume of the unconverted photons is $0.8$ m$ < r<1.93$ m.

\subsection{Displaced dark mediator decay signal}
%We start from the probability density for the decay of light particles when $S$ is produced at rest. 
%This probability for the it can be written as follows:
%\be
%dP_r&=& dW {1 \over L_{r,1}} e^{-r_1/L_{r,1}} {1 \over L_{r,2}} e^{-r_2/L_{r,2}} dr_1 dr_2,
%\label{dw}
%\ee
%with
%\beq
%dW =  {1 \over 4\pi p_0^2} d^3p_{r,1} \delta(|\vec{p}_{r,1}|-p_0),
%\eeq
%where 
%\be
%L_{r,1}=L_{r,2}= p_0 {\tau_{A'} \over m_{A'}},
%\ee
%are the decay lengths of the dark mediators 1 and 2. The subscript ``$r$'' indicates observables taken in the rest frame of $S$. $\tau_{A'}$ and $m_{A'}$ are the proper lifetime and rest mass of the mediator, respectively. $r_i$ is the radial component of 
%the dark mediator $i$. $p_0$  denotes the magnitude of the momentum of the dark mediators in the rest frame of the scalar S. 

In order to obtain a  more realistic evaluation of $P_{\text{acc}}$ than the one given in Eq.~(\ref{acc_naive}), we need to take into account the 
distribution of the initial momentum of heavy resonances ($S$ or $Z'$) affecting the boosts of emerging light particles, which in 
turns translates into a distribution of the decay lengths $L_{A'}$ or $L_{s,a}$. 

Different production mechanisms for $S$ and $Z'$ suggest differences in their boost factors.  
The scalar  $S$ is produced through gluon fusion, which means that the initial states are similarly distributed. 
On the other hand, in our second example, $Z'$ is produced through $q\bar{q}$
initial states, which is asymmetric because it is more probable to find a quark than an anti-quark in a proton due
to differences in their parton distribution functions. As a consequence it is more likely that $Z'$ will have more of  a longitudinal 
boost compared to $S$, while for the latter we find that the production-near-rest picture largely holds. 

Suppose that the distribution of a heavy resonance initial velocities, or boosts, is given by $f(\beta)$. The function satisfies normalization condition \beq \int_{-1}^{1} f(\beta) d \beta =1. \eeq 
We simulate $f(\beta)$ using standard MC tools in practice. Furthermore, given the 
geometry of the detector is cylindrical, and that all decays of light particles to collimated $e^+e^-$ pairs
within radial segments (distance from the origin) $ r_{min}(\theta) < r < r_{max}(\theta)$ pass the photon selection criteria, 
$P_{\text{acc}}$ is proportional to 
\beq
P_{\text{acc}} \propto P_{\rm fid} \times ({\rm Br}_{e^+e^-})^2  ,
\label{eq:problab}
\eeq
where $P_{\rm fid}$ is the probability for dark mediators decaying inside the fiducial regions. $P_{\rm fid}$ can be expressed as
\be
\label{eq:probfid}
P_{\rm fid}&=&\int_{-1}^1 d\beta f(\beta) \int_{\theta_{1,min}}^{\theta_{1,max}} {d\cos\theta_1} \f{\gamma}{2} (1-\beta \cos\theta_1) \left[\sin^2\theta_1 + \gamma^2 (\cos\theta_1 - \beta)^2\right]^{-3/2} \nonumber\\
&&\times \left(e^{-r_{1,min}/L_{L,1}}-e^{-r_{1,max}/L_{L,1}}\right)\left(e^{-r_{2,min}/L_{L,2}}-e^{-r_{2,max}/L_{L,2}}\right),
\ee
with
\be
L_{L,1} &=& p_{L,1} {\tau_{A'}\over m_{A'}} ={p_0 \over \sqrt{1-\cos^2\theta_1 + \gamma^2 (\cos\theta_1 -\beta)^2} }{\tau_{A'}\over m_{A'}},\\
L_{L,2} &=& p_{L,2} {\tau_{A'}\over m_{A'}} = {p_0 \over \sqrt{1-\cos^2\theta_1 + \gamma^2 (\cos\theta_1 -\beta)^2} }\sqrt{1-\cos^2\theta_1 \over 1-\cos^2\theta_2}\; {\tau_{A'}\over m_{A'}}
\ee
are the decay lengths of the dark mediators 1 and 2 in the laboratory frame (denoted with subscript ``$L$"). $\theta_1$ and $\theta_2$ are the polar angles of the dark mediators 1 and 2, respectively in the laboratory frame. $r_{i,min}$ and $r_{i,max}$ are lower and upper boundaries of $r_i$ of the fiducial volume, which both are functions of $\theta$. $\theta_{1,min}$ and $\theta_{1,max}$ represent lower and upper boundaries of $\theta_1$ of the fiducial region. Note that $\theta_1$ and $\theta_2$ are not independent. $\cos\theta_2$ can be expressed in terms of $\cos\theta_1$ and $\beta$
\beq
\cos\theta_2 = \cos\theta_2 (\cos \theta_1, \beta)= -\frac{\beta^2 \cos\theta_1 -2 \beta+\cos\theta_1}{\beta^2-2 \beta \cos\theta_1+1},
\eeq
where $\beta$ is the velocity of the parent particle ($Z'$ or $S$) after the production with a 
boost factor $\gamma \equiv {1 / \sqrt{1- \beta^2}}$. %The subscript ``$L$'' denotes observables in the laboratory frame. 
%$f(\beta)$ is the normalized velocity distribution  of the parent particle, {\em i.e.} $\int d\beta f(\beta) =1$.
We refer the readers to Appendix~\ref{sec:boost} for the more detailed derivation of the decay probability including the boost effect.

Given the geometry of the detector and the probability (\ref{problab}), we can calculate $P_{\rm fid}$ for both 750 GeV scalar and vector resonance scenarios. 
We give the results for $P_{\text{fid}}$ for the 750 GeV scalar resonance scenario in Tab.~\ref{tab:fr} as an example. $P_{\rm fid}$ a function of decay length $L_{A'}$. One can observe that as the decay length grows, the $P_{\rm fid}$ drops precipitously.

\begin{table}[htbp]
   \centering
\begin{tabular}{c c c c c }
\hline
 Decay Length ($L_d$) & \text{Converted} & Unconverted 1 & Converted 1+2 &  Unconverted 1+2 \\
 \hline 
 0.1 & $0.31$ & $2.9 \times 10^{-8}$ & $0.32$ & $2.9 \times 10^{-8}$   \\
 1 & 0.38 & $3.4\times 10^{-2}$ & 0.42 & $5.0\times 10^{-2}$ \\
 10 & $1.5\times 10^{-2}$ & $3.9\times 10^{-3}$ & $1.8\times 10^{-2}$ & $7.4\times 10^{-3}$ \\
 20 & $4.0\times 10^{-3}$ & $1.1\times 10^{-3}$ & $5.0 \times 10^{-3}$ & $2.2\times 10^{-3}$ \\
 100 & $1.7\times 10^{-4}$ & $5.1\times 10^{-5}$ & $2.2\times 10^{-4}$ & $1.0\times10^{-4}$ \\
\hline
\end{tabular}
 \caption{Probabilities of dark photon decays inside the ATLAS detector, $P_{\rm fid}$, for the 750 GeV scalar resonance scenario. Various decay length $L_d$ and fiducial regions are considered. Events with at least one of the decays occurring inside the tracker volume are categorized as ``Converted''. The ``Unconverted 1" category includes events where both dark mediators decay inside the remaining part of the fiducial volume (gap region and the first layer of the ECAL). Similarly, the ``Converted 1+2" and ``Unconverted 1+2" categories are the generalization of the Converted and Unconverted 1 categories by including the second layer of the ECAL into the fiducial volume of the event reconstruction.}
 % events where both of the dark mediators decay inside the first and second layers of the EM calorimeter. Converted1+2 category includes events where at %least one of the dark mediators decaay occurs inside the tracker while the other can be in }
   \label{tab:fr}
\end{table}

From \eqref{problab} to obtain the final $P_\text{acc}$, we still need to multiple the right hand side by the acceptance rate and diphoton reconstruction efficiency, {\em i.e.},
\beq
P_\text{acc} = \epsilon^2_{\gamma} \times A \times P_{\rm fid} \times ({\rm Br}_{e^+e^-})^2,
\label{eq:pacc}
\eeq 
where $\epsilon_\gamma = 95\%$ is the reconstruction efficiency for a single photon~\cite{ATLAS:2015xxx}.  The selection cuts on $|\eta|$ has already been considered in the calculation $P_\text{fid}$. The rest selection cuts in~\cite{ATLAS:2015xxx} are as follows:
\beq
E_{\rm T}^{\gamma_1} > 40 \gev, \quad E_{\rm T}^{\gamma_2} > 30 \gev, \quad E_{\rm T}^{\gamma_1}/m_{\gamma\gamma} >0.4, \quad E_{\rm T}^{\gamma_2}/m_{\gamma\gamma} >0.3.
\eeq 
We use Monte-Carlo simulation to implement above cuts and obtain the acceptance $A$. The resulting acceptance $A$ (after $|\eta|$ cuts) is $68\%$ ($84\%$) for 750 GeV scalar (vector) resonance scenario. Substituting the acceptance and efficiency back to \eqref{pacc}, we get $P_\text{acc}$ that consequently yields $\sigma_\text{Signal}$ through \eqref{mformula}.

\subsection{Preferred region of light particle parameter space}

\begin{figure}[t]
   \includegraphics[width=0.48\textwidth]{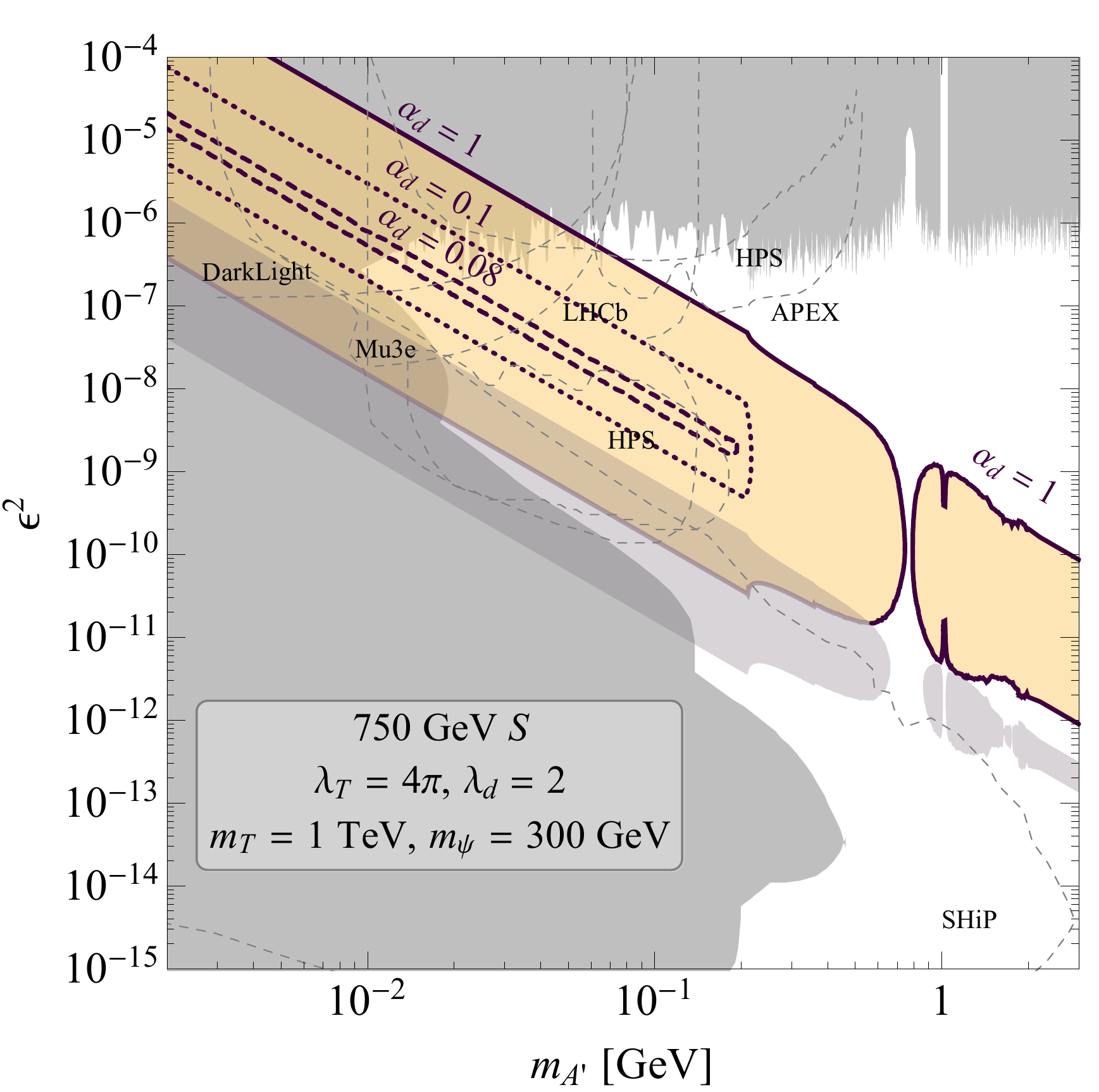} % requires the graphicx package
      \includegraphics[width=0.48\textwidth]{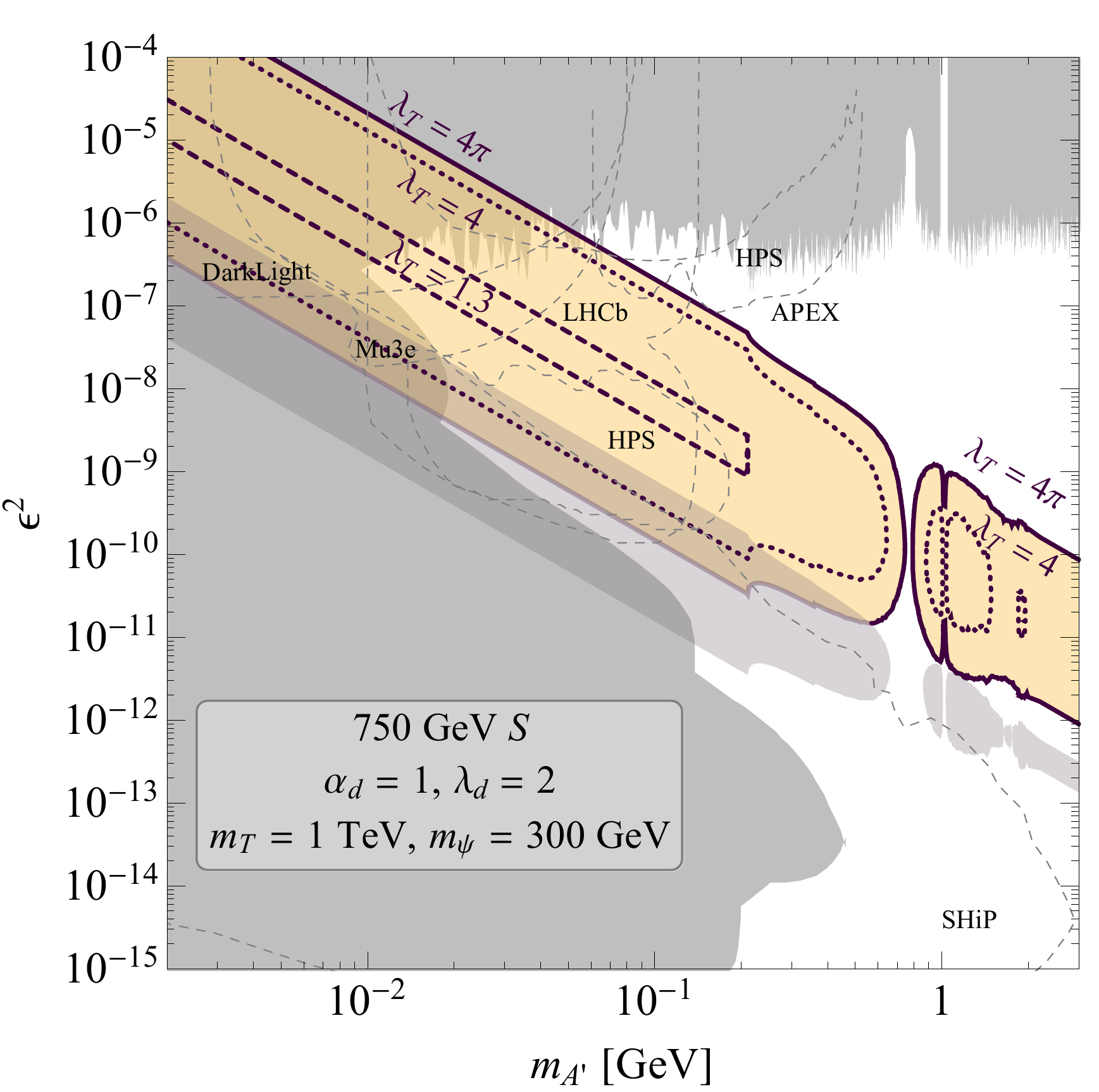} % requires the graphicx package
      
   \caption{Preferred parameter space (yellow shaded) in the $\epsilon^2$ versus $m_{A'}$ plane for dark photons that can explain the 750 GeV scalar resonance through faking photons. In the left and right panel, we vary the parameter $\alpha_d$ and $\lambda_T$ separately. The solid, dotted, and dashed lines in the left(right) panel respectively represent parameters corresponding to 30 observed diphoton events for $\alpha_d=1$, 0.01, and 0.08 ($\lambda_T = 4\pi$, 4, and 1.3) with a fixed $\lambda_T =4\pi$ ($\alpha_d=1$). Other parameters in the calculation are set to be $\lambda_d = 2$, $m_\psi = 300$~GeV and $m_T = 1 ~\text{TeV}$. The purple-gray shaded regions are excluded by the mono-photon search at the ATLAS~\cite{Aad:2014tda}. It excludes part of parameter space for $\alpha_d=1,\lambda_T=4\pi$ that we marked as purple-gray lines. Nevertheless, the mono-photon search does not further exclude preferred parameter space for smaller $\alpha_d$ and $\lambda_T$ values listed in the plot. In the plot, we also include current constraints and future prospects on the $\epsilon^2$ versus $m_{A'}$ plane for dark photons that decay directly to SM particles (see {\em e.g.}~\cite{Alekhin:2015byh} and reference in \S I).}
   \label{fig:daplimit}
\end{figure}

In this subsection we perform a ``fusion" of all different components of our calculation in order to derive the 
allowed parameter space for light particles. Our strategy is to be conservative, which means we should allow the 
largest possible variations in the properties of the 750~GeV resonance. To that effect, we take the largest possible range for the 
coupling that regulates the production of $S$ through the gluon fusion, $0\leq \lambda_T \leq 4 \pi$. The upper boundary 
would correspond to the largest production cross section, and therefore admits the lowest possible $P_{\text{acc}}$. 
At this point we will also assume that 
every electron-positron decay of light particles is going to pass the photon selection criteria.  
Violation of this assumption in practice is possible for higher $A'$ masses, which would reduce 
the region of interest on the $\epsilon-m_{A'}$ parameter space.

A fixed minimum value for the $P_{\text{acc}}$ has, of course, two solutions in terms of $L_{A'}$. If the decay length is too short, all the 
decays will happen inside or close to the beam pipe, while if the decay length is too large, only a small finite number of 
$A'$ pairs would decay in or before the ECAL. 
%We present results in the models discussed in Sec.~\ref{sec:TheoreticalMotivation} using the approach in Sec.~\ref{sec:prob}.
For the dark photon model, we obtain the allowed region that would be consistent with our scenario  for the 
750~GeV resonance. The preferred part of the dark photon parameter space is shown in Fig.~\ref{fig:daplimit} with 
with $m_T = 1~{\rm TeV},~m_\psi=300~{\rm GeV}, ~\lambda_d = 2$ while $\lambda_T $ and $\alpha_d$ are varied. 
(Notice that this choice of $m_\psi$ and $\alpha_d$ fits the reported width of a possible 750~GeV resonance). 
The yellow shaded region is favored by the 750~GeV resonance.
One can also see that the allowed parameter space has a band structure, which follows from the 
$L_{A'} \propto (\epsilon m_{A'})^{-2}$ scaling. Wiggles, deviations and a dip near 1 GeV occurs due to the enhancement of hadronic 
decays of  ${A'}$ and the reduction of ${\rm Br}_{e^+e^-}$. In the left panel of the plot $\lambda_T$ is set to its maximum value while $\alpha_d$ is varied, while on the right panel 
$\alpha_d=1$ and $\lambda_T$ is scanned. We observe that as the couplings diminish so does the allowed part of the 
parameter space. However, some allowed parameter space still exists for $\lambda_T \sim O(1)$ or $\alpha_d \sim O(0.1)$. 
It is also worth mentioning that above $m_{A'} = 2 m_\mu$ there is an appreciable branching to muons, 
so that one should expect ``fake photon" and muon pair, or two muon pair events appearing in the same model 
that should reconstruct to the same invariant mass. 

On the whole, one can see that intensity frontier searches cannot fully exclude the suggested region of the model parameter space. 
It is easy to understand why: in the adopted LHC scenario, $A'$ particles have relatively small mixing angles 
$\epsilon \sim \mathcal O(10^{-4})$, which for most fixed target searches would not lead to detectable displaced decays. 
At the same time, it is too small a coupling to be currently  ruled out by the search for ``bumps'' in the $e^+e^-$ spectrum. 
 We also include the exclusion region imposed  by the ATLAS 
mono-photon constraints~\cite{Aad:2014tda}. This constraint  comes from the situation when one $A'$ decays 
before or inside the ECAL faking a photon, while the second $A'$ completely escapes the detector before decaying. 
The current limit on the cross section is 6.1 (5.3) fb at 95\% C.L. This constraint will be relevant for the 
longer $L_{A'}$, and this is seen on the left of Fig. \ref{fig:daplimit} with the gray band being parallel but below the yellow one. 
It is worth mentioning that future intensity frontier experiments can potentially exclude some part of preferred parameter space in 
Fig~\ref{fig:daplimit}. These projected limits are shown in dashed lines in Fig~\ref{fig:daplimit}.
In addition, with more data collected the mono-photon search should be able to provide a stronger constraint at the LHC run-II.
%To translate the limit to 
%$\epsilon^2-m_{A'}$ plane we assume that one of the dark photon decays inside the ATLAS detector while the other escapes. 

\begin{figure}[t]
   \includegraphics[width=0.5\textwidth]{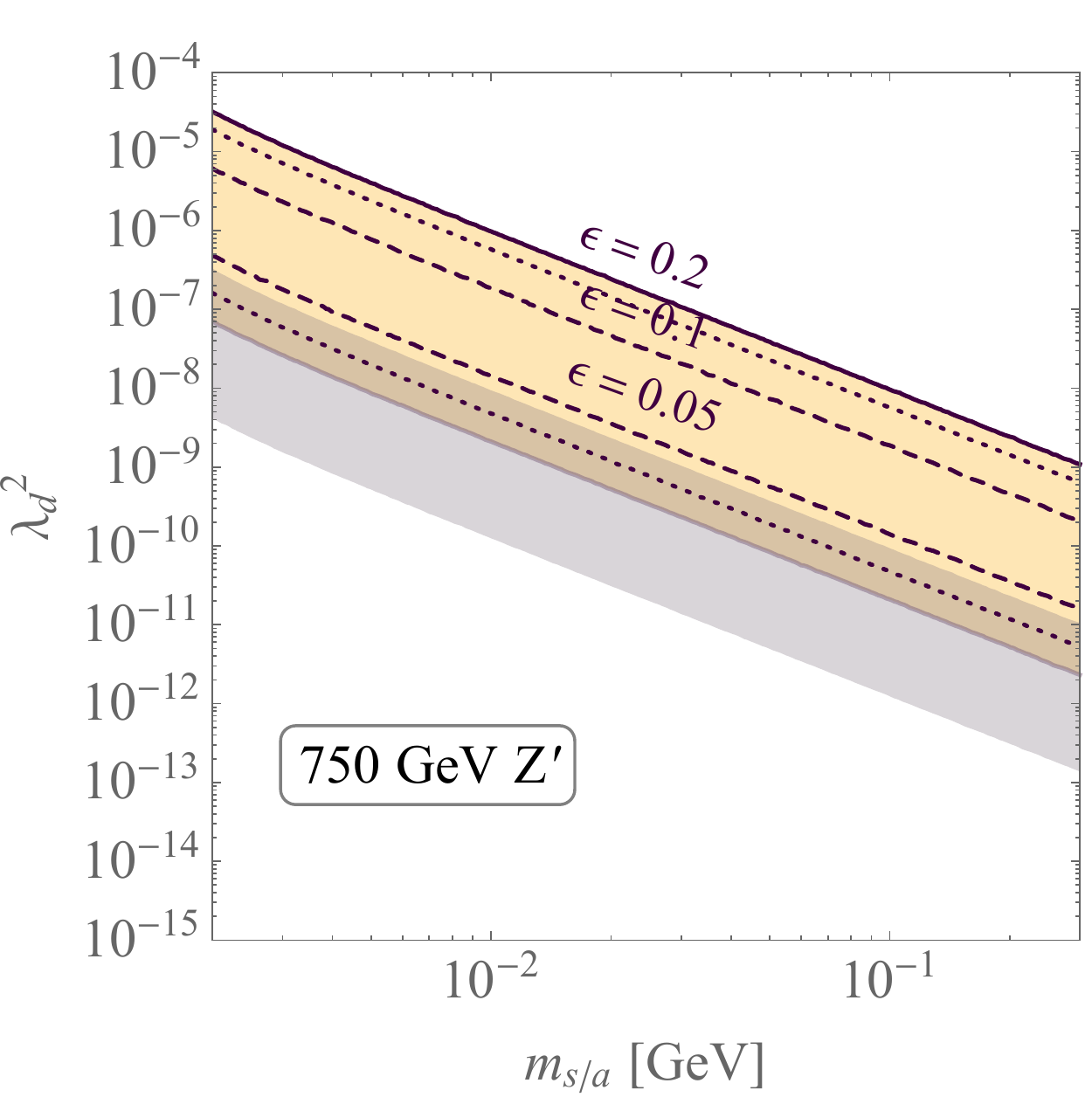} % requires the graphicx package   
      \caption{Preferred parameter space (yellow shaded) in the $\lambda_d^2$ versus $m_{s/a}$ plane for dark scalars that can explain the 750 GeV vector resonance through faking photons. The solid, dotted, and dashed lines represent parameters corresponding to 30 observed diphoton events for $\epsilon=0.2$, 0.1, and 0.05, respectively. The purple-gray shaded regions are excluded by the mono-photon search at ATLAS~\cite{Aad:2014tda}. It excludes part of parameter space for $\epsilon=0.2$ that we marked as a purple-gray line.  The mono-photon search does not further exclude preferred parameter regions for $\epsilon = 0.1$ and 0.05.}

   \label{fig:zplimit}
\end{figure}

Next we present the result of the $Z'$ model in the $\lambda_{d}^2-m_{s/a}$ parameter space in Fig.~\ref{fig:zplimit} with various
 contours corresponding to $\epsilon=0.05, 0.1$ and $0.2$. The yellow shaded region are favored by the 750 GeV $Z'$ resonance while the gray shaded region is excluded by the mono-photon searches with $\epsilon=0.2$. For $m_{s/a}=0.1$ GeV
 one can have $\lambda_d^2$ between $10^{-10}$ and $10^{-8}$. The allowed range of $\epsilon$ in the yellow shaded region 
 is $0.02 \lesssim \epsilon \lesssim 0.2$. The lower limit is to ensure having enough production cross section while the upper limit comes from  the electroweak precision test~\cite{Hook:2010tw}. Notice that the model 
with $\epsilon<0.1$ and $g_d \gg g_1 \epsilon$ is very difficult to constrain via ``conventional" $q\bar q \to Z' \to \mu^+\mu^-$
searches due to a small branching ratio for the $Z'$ decay to SM particles, which leads to $\epsilon^4$ scaling of the signal. 

\section{Discussion}
\label{sec:Discussion}
\subsection{Potential methods to exclude models with dark mediators}
So far there is only limited amount of data available. However, with more data it is likely that one can 
statistically discriminate between  real photon events and decays of dark mediators. While the 
properties of the photons are of course fully specified by QED and atomic physics, the main input 
parameters for the dark mediator decays will be its energy, mass and the decay length (such as $E_{A'},~m_{A'}$ and $L_{A'}$ 
as in the dark photon example). Below we outline 
important differences between the conversions of real photons and decays of dark mediators. 

\begin{enumerate}

\item  {\em Affinity of conversions to the material inside the detector.} Photons convert to pairs in the field of the nucleus, 
and therefore the distribution of conversion points roughly follows the number density of atoms weighted with the square of the 
atomic number, $Z^2n_A$. The dark mediators, on the other hand, can decay anywhere in the detector, including hollow parts.
The distribution of vertices for the converted photon events should provide a useful discrimination.

\item {\em Events beyond the first layer of the ECAL.}
The decays of dark mediators can occur in the ECAL {\em beyond} the first layer of the calorimeter,
which would correspond to an unusual penetration depth for a regular photon. In fairness, 
the probability of decay within the second or third layer of the calorimeter is not very large for the 
models considered, and more data is needed for this criterion to become useful. But even with current statistics, 
the searches of ``late converting" photons in association with regular photons is of interest and should be 
pursued.

\item {\em Distribution of converted vs unconverted events.} Exponential dependence on the distance travelled, 
for a short decay length $L_d$, will always enhance the fraction of fake unconverted events. That is most dark mediators
at short $L_d$ will decay before reaching the ECAL. Therefore this can be a useful criterion for part of the parameter space. 

\item {\em Energy distribution of electron-positron pairs.}
It is well known that the electron-positron pairs created by Bethe-Heitler process (regular conversion) 
have an appreciable fraction of events with asymmetric energy distribution ($E_{e^+} \gg E_{e^-} $ or $ E_{e^-} \gg E_{e^+} $)
whereas a vast fraction of dark mediator decays has $E_{e^+} \sim E_{e^-} $. This fact is well appreciated in the direct 
dark photon searches. An abnormally low fraction of asymmetric pairs could be a signature of dark mediators. 

\item {\em Shape and point of origin for the shower.}  Unconverted photons may have a small but non-zero penetration depth inside the 
first layer of the ECAL, while dark photons decaying in the gap between the tracker and the ECAL enter the calorimeter as pairs, and thus 
shower immediately. This will affect the shape of the shower, its starting point, and possibly the energy reconstructed from the 
standard procedures. 

\item {\em Abnormal separation of electron-positron pair.}  In this paper we have avoided the discussion of 
the drop in efficiency for converted and/or unconverted photons when the mass of the dark mediator become large.
When a dark mediator such as $A'$ decays to the electron-positron pair, each electron receives a perpendicular momentum 
$p_\perp \sim m_{A'}/E_{A'}$. After some distance travelled, this may lead to an abnormally large separation of electrons and  positrons,
compared to a similar behavior of a regular conversion pair, when they cross a layer of the pixel detector and/or reach the ECAL. 
Detailed implementation of this criterion should determine the {\em maximum} mass for a dark mediator capable of 
faking a photon. 

\end{enumerate} 

We believe that the possibility of dark mediators mimicking real photons deserves a closer 
look by the experimental collaborations. A few items outlined above may serve as a basis for developing a 
statistical procedure that would emphasize or suppress fake photons vs real photons and vice versa. 
It is also worth mentioning that due to the difference in the linear sizes of the ATLAS and CMS detectors
(hence a different sensitivity to $L_d$), there can be an additional discriminating power in a combined treatment. 
Also, it may be that dark mediator decays create a large number of events that are neglected for one or many of the 
above reasons. Therefore a closer look in a sample with loosened 
criteria for photon identification may also contribute to constraining or validating dark mediator models.

%\begin{figure}[b]
%\vspace{0.5cm}
%\centerline{\includegraphics[width=0.36\columnwidth]{bg1.pdf}
%\includegraphics[width=0.25\columnwidth]{bg2.pdf}
%\includegraphics[width=0.38\columnwidth]{bg3.pdf}} 
%\caption{The dominant backgrounds are (a) offshell decay of the SM photon into dimuon, (b) Bethe-Heitler trident reaction and (c) vector %mesons decay into dimuon.} 
%\label{fig:bg}
%\end{figure}

\section{Conclusion}
\label{sec:Conclusion}

In this paper we have considered the exotic possibility that metastable BSM particles of low mass could be produced as a result of 
a heavy resonance decay. Being weakly unstable, these particles decay 
to electron-positron pairs that may in fact resemble the conversion pairs originating from a regular photon. 
The prime candidates for such metastable particles are dark photons, as well as light scalars and pseudoscalars, which all have 
small branchings to neutrinos and therefore do not generate a large missing transverse momentum signal. We have examined 
both possibilities, without imposing very restrictive assumptions on the properties of the 750~GeV resonance. 
We have found that the parameter space for light particles ({\em e.g.} the dark photon models prefer a
somewhat wide range of parameters along the 
$m_{A'}/(100~{\rm MeV})\times (\epsilon/10^{-4}) \sim O(1)$ line) that emerges from this analysis is not
excluded by the current limits. However, a number of new proposals at different stages of maturity exists 
\cite{Stepanyan:2013tma,Balewski:2014pxa,Echenard:2014lma,Ilten:2015hya,Abe:2010gxa} which will eventually probe deep inside the region of interest. 

In the models we consider, the mono-photon searches provide an important constraint. Also, should the light particles be able to decay to muons, 
a search of two collimated muons plus a ``fake photon" reconstructing to the same invariant mass is a 
promising search channel.  

Irrespective of the future status of the 750~GeV resonance, it seems important for the experimental collaborators to build statistical 
discriminators that would allow (given enough data) to distinguish between regular SM photon events and 
would-be-photon dark mediator decays. We have provided a discussion of some avenues along which this problem 
might be addressed.

\vspace{2cm}

{\it Note added-}

As this paper was being finalized, we became aware of recent preprints \cite{Bi:2016gca,wang}
that also have detailed discussions on the possibility that the diphoton resonance  may 
also be explained by dark mediators mimicking real photon signals,
and thus have overlap with some of the discussions in our paper.

\section*{Acknowledgements}

We would like to thank A. Soffer and Y. Zhang for helpful discussions. Y.Z. would like to thank R. Essig for the discussion of $Z'$ models.
The work of C.-Y.C, M.L. and M.P. is supported by NSERC, Canada. 
Research at the Perimeter Institute is supported in part by the Government of Canada through NSERC and by the Province of Ontario through MEDT. Y.Z. is supported through DoE grant DESC0008061.

\appendix

\section{Kinetic mixing}
\label{km}
The relevant Lagrangian for the kinetic mixing includes three parts: the abelian gauge part, $\mathcal L_{\text{gauge}}$, the vector boson mass part, $\mathcal L_{\text{mass}}$, and the interactions between gauge bosons and SM fermions, $\mathcal L_{\text{int}}$. The abelian gauge part is given by 

\beq
\mathcal L_{\text{gauge}}= -\frac{1}{4}B^2_{\mu\nu}-\frac{1}{4}{F'}_{\mu \nu}^2+\frac{\sin \epsilon_Y}{2}{F'}_{\mu\nu}B^{\mu\nu},
\eeq

where $B_{\mu\nu}=\partial_{[\mu} B_{\nu]}$ and $F'_{\mu\nu}=\partial_{[\mu} A'_{\nu]}$. Here we use $\sin \epsilon_Y$ as the kinetic mixing coefficient to simplify later expressions. Note that $\lim_{\epsilon_Y\to 0}\sin \epsilon_Y=\epsilon_Y$. The mixing term $F' B$ is removed by field redefinition

\beq
\begin{pmatrix}
\tilde B_\mu\\
\tilde A'_\mu
\end{pmatrix}
=
\begin{pmatrix}
1 & -\sin \epsilon_Y\\
0 & \cos \epsilon_Y
\end{pmatrix}
\begin{pmatrix}
 B_\mu\\
 A'_\mu
\end{pmatrix},
\eeq
where the fields with tilde are the redefined fields. We follow the standard  symmetry breaking conventions for the field $B$. 
The SM $SU(2)\times U(1)$ covariant derivative can be written as
\beq
D_\mu=\partial_\mu - ig W^a_\mu T^a -i g' Y  B_\mu,
\eeq
and rotated into
\beq
D_\mu=\partial_\mu-i\frac{g}{\sqrt 2} \left(W^+_\mu T^+ +W^-_\mu T^-\right)-i\frac{g}{\cos \theta_W} Z_\mu (T^3-\sin^2 \theta_W Q)-i e  A_\mu Q,
\eeq
where $\theta_W$ is the weak mixing angle, $g$ and $g'$ are SM $SU(2)_L$ and $U(1)_Y$ gauge couplings. $SU(2)_L$ charge $T^3$, $U(1)_Y$ charge $Y$, and $U(1)_{EM}$ charge $Q$ are related by $Q\equiv T^3+ Y$. Rotating the redefined neutral vector boson fields yields $\tilde A_\mu$, $\tilde Z_\mu$ and $\tilde Z'_\mu$,
\beq
\begin{pmatrix}
\tilde A_\mu\\
\tilde Z_\mu\\
\tilde Z'_\mu
\end{pmatrix}
=
\begin{pmatrix}
c_W & s_W & 0\\
-s_W & c_W & 0\\
0 & 0 & 1
\end{pmatrix}
\begin{pmatrix}
\tilde B_\mu\\
 W^3_\mu\\
\tilde A'_\mu
\end{pmatrix}
=
\begin{pmatrix}
c_W & s_W & -c_W s_\epsilon\\
-s_W & c_W & s_W s_\epsilon\\
0 & 0 & c_\epsilon
\end{pmatrix}
\begin{pmatrix}
 B_\mu\\
 W^3_\mu\\
 A'_\mu
\end{pmatrix}
\eeq
Above we begin to use a short-handed notation, {\em i.e.}, $c$, $s$, $t$ stands for $\sin$, $\cos$, $\tan$ respectively.  The subscript stands for the function variables and $W$ stands for $\theta_W$.

The relevant mass part of the Lagrangian is
\bea
\mathcal L_{\text{mass}}={}&\frac{1}{2} m_{A'}^2 {A'}_\mu^2+\frac{1}{2} m_{Z^0}^2 \begin{pmatrix}
B_\mu & W^3_\mu 
\end{pmatrix}
\begin{pmatrix}
s_W^2 & -c_W s_W \\
-c_W s_W & c_W^2 \\
\end{pmatrix}
\begin{pmatrix}
B^\mu \\ W^{3, \mu}
\end{pmatrix}\\
={}&\frac{1}{2} m_{A'}^2 {\tilde {Z'}}_\mu^2/c_\epsilon^2+ \frac{1}{2} m_{Z^0}^2 \begin{pmatrix}
\tilde A_\mu & \tilde Z_\mu & \tilde Z'_\mu 
\end{pmatrix}
\begin{pmatrix}
0 & 0 & 0\\
0 & 1 &  -s_W t_\epsilon \\
0 & -s_W t_\epsilon & s_W^2 t_\epsilon^2 \\
\end{pmatrix}
\begin{pmatrix}
\tilde A^\mu\\
\tilde Z^\mu\\
\tilde {Z'}^\mu
\end{pmatrix}\\
={}& \frac{1}{2}  \begin{pmatrix}
\tilde A_\mu & \tilde Z_\mu & \tilde Z'_\mu 
\end{pmatrix}
\begin{pmatrix}
0 & 0 & 0\\
0 & m_{Z^0}^2 & -m_{Z^0}^2 s_W t_\epsilon \\
0 & -m_{Z^0}^2 s_W t_\epsilon & m_{Z^0}^2 s_W^2 t_\epsilon^2 +m_{A'}^2/c_\epsilon^2\\
\end{pmatrix}
\begin{pmatrix}
\tilde A^\mu\\
\tilde Z^\mu\\
\tilde {Z'}^\mu
\end{pmatrix}
\label{eq:dep}
\eea
In the last step of the above equation, the Lagrangian is expressed in the redefined fields. However, the mass matrix is still non-diagonal. We introduce one more rotation to eliminate the mixing,  
\bea
 \mathcal L_{\text{mass}}={}& \frac{1}{2}  \begin{pmatrix}
\tilde A_\mu & \tilde Z_\mu & \tilde Z'_\mu 
\end{pmatrix}
\begin{pmatrix}
1 & 0 & 0\\
0 & c_\xi & -s_\xi \\
0 & s_\xi & c_\xi
\end{pmatrix}
\begin{pmatrix}
0 & 0 & 0\\
0 & m_{\bar Z}^2 & 0 \\
0 &0 & m_{\bar Z'}^2
\end{pmatrix}
\begin{pmatrix}
1 & 0 & 0\\
0 & c_\xi & s_\xi \\
0 & -s_\xi & c_\xi
\end{pmatrix}
\begin{pmatrix}
\tilde A^\mu\\
\tilde Z^\mu\\
\tilde {Z'}^\mu
\end{pmatrix}\\
\equiv{}&\frac{1}{2}  \begin{pmatrix}
\bar A_\mu & \bar Z_\mu & \bar Z'_\mu 
\end{pmatrix}
\begin{pmatrix}
0 & 0 & 0\\
0 & m_{\bar Z}^2 & 0 \\
0 &0 & m_{\bar Z'}^2
\end{pmatrix}
\begin{pmatrix}
\bar A^\mu\\
\bar Z^\mu\\
\bar {Z'}^\mu
\end{pmatrix}
 \eea
We use the bar to represent fields in their physical (mass) basis. The eigenstates of the mass matrix in (\ref{eq:dep}) yield $m_Z^2$ and $m_{Z'}^2$ as 
\beq
m_{\bar Z, \bar Z'}^2=\frac{m_{Z^0}^2}{2}\left[\cos^2 \theta_W+\sec^2 \epsilon_Y \sin^2 \theta_W+\delta^2 \pm\text{sign}\left(1-\delta\right)\sec^2 \epsilon_Y \sqrt{\Xi}\right],
\eeq
where 
\beq
\Xi\equiv-4\delta^2 \cos^4 \epsilon_Y+\left[\left(1+\delta^2\right)\cos^2 \epsilon_Y + \sin^2\epsilon_Y \sin^2 \theta_W\right]^2
\eeq 
and
\beq
\delta^2 \equiv \frac{m_{A'}^2}{m_{Z^0}^2}.
\eeq
$m_{Z^0}$ is the SM $Z$-boson mass before kinetic mixing. The appearance of the sign function is because we intend to assign $m_Z=m_{Z^0}$ when the mixing coefficient vanishes.

Comparing the mass matrix before and after the diagonalization, we found  
\begin{align}
\tan \xi ={}&\frac{m_{Z^0}^2\sin \theta_W \tan \epsilon_Y}{m_{Z'}^2-m_{Z^0}^2}\\
={}&\frac{ 1-\delta ^2 -\sin^2 \theta_W \tan^2 \epsilon_Y -\text{sign}(1-\delta)  \sec^2 \epsilon_Y \sqrt{\Xi}}{2 \tan \epsilon_Y \sin \theta_W}.
\label{eq:angle}
\end{align}

We have adopted multiple matrix transformations so far. The transfer matrix from the original gauge basis to the final physical basis is

\beq
\begin{pmatrix}
\bar A_\mu\\
\bar Z_\mu\\
\bar {Z'}_\mu
\end{pmatrix}
=\begin{pmatrix}
c_W & s_W & -c_W s_\epsilon\\
-c_\xi s_W & c_\xi c_W & c_\epsilon s_\xi+c_\xi s_\epsilon s_W \\
s_\xi s_W & -s_\xi c_W &  c_\epsilon c_\xi -s_\epsilon s_\xi s_W
\end{pmatrix}
\begin{pmatrix}
 B_\mu\\
 W^3_\mu\\
 A'_\mu
\end{pmatrix}.
\eeq

The interaction part between the gauge bosons and the SM fermions can be written as
\beq
\mathcal L_{\text{int}}=e A^\mu J_{\text{EM},\, \mu}+g Z^{0,\, \mu} J_{Z^0,\, \mu}
\eeq
where $e$ is the SM $U(1)_{EM}$ gauge coupling, and $J_\text{EM}$ and $J_{Z^0}$ are the SM EM and neutral current, respectively. $A$ and $Z^0$ can be expressed in the physical basis as
\beq
\begin{pmatrix}
 A_\mu\\
 Z^0_\mu
 \end{pmatrix}
=\begin{pmatrix}
c_W & s_W & 0  \\
-s_W & c_W & 0
\end{pmatrix}
\begin{pmatrix}
 B_\mu\\
 W^3_\mu\\
 A'_\mu
\end{pmatrix}
=
\begin{pmatrix}
1 & c_W  t_\epsilon s_\xi & c_W t_\epsilon c_\xi \\
0 & -s_W t_\epsilon s_\xi+c_\xi & -s_W t_\epsilon c_\xi - s_\xi
\end{pmatrix}
\begin{pmatrix}
\bar A_\mu\\
\bar Z_\mu \\
\bar Z'_\mu
\end{pmatrix}.
\eeq
Therefore the interaction in the physical basis is
\beq
\mathcal L_{\text{int}}=\begin{pmatrix}
 \bar A_\mu &
  \bar Z_\mu &
  \bar Z'_\mu
\end{pmatrix}
\begin{pmatrix}
1 & 0 \\
c_W  t_\epsilon s_\xi & -s_W t_\epsilon s_\xi+c_\xi  \\
c_W t_\epsilon c_\xi & -s_W t_\epsilon c_\xi -s_\xi\\
\end{pmatrix}
\begin{pmatrix}
e J_{\text{EM}}^\mu\\
g J_{Z^0}^\mu\\
\end{pmatrix}.
\eeq

From the above expression, we read off the relevant part for the  $Z' \bar f f$ interaction as
\beq
{\mathcal L}_{\bar Z' \bar f f} =\frac{g}{\cos \theta_W} \left[-\sin \xi \left(T^3 \cos^2 \theta_W- Y \sin^2 \theta_W\right)+\tan \epsilon_Y \cos \xi \sin \theta_W Y\right] \bar Z'_\mu \bar f \gamma^\mu f.
\eeq
Similarly, the $Z \bar f f$ interaction is modified to
\beq
{\mathcal L}_{\bar Z \bar f f} =\frac{g}{\cos \theta_W} \left[\cos \xi \left(T^3 \cos^2 \theta_W- Y \sin^2 \theta_W\right)+\tan \epsilon_Y \sin \xi \sin \theta_W Y\right] \bar Z_\mu \bar f \gamma^\mu f.
\eeq

In the limit $m_{A'} \ll m_{Z^0}$ and $\epsilon_Y \ll 1$, we have
\beq
{\mathcal L}_{\bar Z' \bar f f} \simeq \epsilon_Y \cos \theta_W e Q \bar Z'_\mu \bar f \gamma^\mu f\equiv \epsilon e Q \bar Z'_\mu \bar f \gamma^\mu f.
\eeq
with
\beq
m_{\bar Z'}^2 \simeq m_{A'}^2 \left(1-\epsilon_Y^2 \sin^2 \theta_W\right).
\eeq
Above we introduce the ``usual" kinetic mixing parameter  $\epsilon\equiv\epsilon_Y\cos \theta_W$. In the other limit $m_{A'} \gg m_{Z^0}$ and $\epsilon_Y \ll 1$, we obtain
\beq
{\mathcal L}_{\bar Z' \bar f f} \simeq= \epsilon_Y g' Y \bar Z'_\mu \bar f \gamma^\mu f= \f{\epsilon_Y}{\cos \theta_W} e Y \bar Z'_\mu \bar f \gamma^\mu f=\f{\epsilon e Y}{\cos^2 \theta_W}  \bar Z'_\mu \bar f \gamma^\mu f
\eeq
with
\beq
m_{\bar Z'}^2 \simeq m_{A'}^2.
\eeq
Note that the couplings between $Z'$ and fermions are independent  of the $Z'$ mass in both limits. 

\section{Decay probability with boost effect}
\label{sec:boost}
In this section, we derive the decay probability of dark mediators in the laboratory frame. We first start with 
 the decay probability in the rest frame of the parent particle $S$ which can be written as
\be
P_r&=&\int dW {1 \over L_{r,1}} e^{-r_1/L_{r,1}} {1 \over L_{r,2}} e^{-r_2/L_{r,2}} dr_1 dr_2, \\
\rm{with}\;\;\;&& dW =  {1 \over 4\pi p_0^2} d^3p_{r,1} \delta(|\vec{p}_{r,1}|-p_0). 
\label{dw}
\ee
The normalization factor in Eq. \ref{dw} is chosen such that $W = 1$ after carrying out the integration.   $p_0 \simeq m_S/2$ is the magnitude of the momentum of the daughter particles and the subscript ``$r$'' indicates that the observable is in the rest frame of the parent particle. $r_i$ is the radial coordinate of 
the dark mediator $i$. We neglect the mass of the daughter particles since they are much lighter than the parent particle. 
The delta function is used to impose the on-shell condition. Note that the momenta of the two daughter particles are related
$\vec{p}_{r,1}=-\vec{p}_{r,2}$ so the delta function requires both daughter particles to be on-shell.
$r_i$ and $L_{r,i}$ are radial coordinate and the decay length of the daughter particle $i$ ($i=1,2$) in the rest frame of the parent particle. We also assume that the boost is only along the beam-pipe, {\em i.e.} the $z$ direction. Using a Lorentz transformation one can express the $z$ component of the momentum of the daughter particle in the rest frame in terms of the observables in the laboratory frame
\be
p_r^z = \gamma (p_L^z - \beta E_L)\simeq \gamma \; p_L( \cos\theta- \beta ), 
\ee
where $p_L^z$ and $E_L$ are the $z$ component momentum and the energy of the daughter particle in the laboratory frame, respectively.
The subscript $L$ indicates that the observable is in the laboratory frame.
$\theta$ represents the polar angle of the daughter particles in the laboratory frame. We have used $p_L^z = p_L \cos\theta$ and $E_L \simeq p_L$ in the last step. 
$\beta$ is the relative velocity between the rest frame of the parent particle and the laboratory frame. The boost factor $\gamma = 1/\sqrt{1-\beta^2}$. 
Similarly we obtain the following equations
\be
dp_r^z &=& \gamma (1 - \beta {p_L^z \over E_L})dp_L^z \simeq \gamma (1 - \beta \cos\theta)dp_L^z, \label{dp}\\
|\vec{p}_r|&=&\sqrt{p_\perp^2 + \gamma^2(p_L^z -\beta E_L)^2 } \simeq p_L\sqrt{\sin^2\theta + \gamma^2 (\cos\theta - \beta)^2},
\label{vp}
\ee
where $p_\perp = p_L \sin\theta$ is the momentum of the daughter particle in the transverse plane. 
In the last steps of Eqs.~\ref{dp} and \ref{vp} we have used the relations, $p_\perp = p_L \sin\theta$ and $p_L^z = p_L \cos\theta$. Therefore the 
$\delta$ function in \ref{dw} in the laboratory frame can be written as
\be
\delta(|\vec{p}_r|-p_0) &=& \delta(p_L\sqrt{\sin^2\theta + \gamma^2 (\cos\theta - \beta)^2}-p_0) \nonumber\\
                      &=& {1 \over \sqrt{\sin^2\theta + \gamma^2 (\cos\theta - \beta)^2}} \delta(p_L-{p_0 \over \sqrt{\sin^2\theta + \gamma^2 (\cos\theta - \beta)^2}}).
\ee
Likewise, $dW$ in the laboratory frame is as follows 
\be
dW &=&{d\cos\theta \over 2} dp_L \delta(p_L - {p_0\over \sqrt{\sin^2\theta + \gamma^2 (\cos\theta - \beta)^2}}) \nonumber\\
&\times& \gamma (1-\beta \cos\theta) (\sin^2\theta + \gamma^2 (\cos\theta - \beta)^2)^{-3/2}.
\ee
Furthermore, based on the momentum conservation we know that there are relations between daughter particles 1 and 2. 
\be
%p_{r,i}^z &=& \gamma ( p_{L,i}^z- \beta E_L)
p_{r,1}^z &=& -p_{r,2}^z, \\
p_{\perp,1} &=& p_{\perp,2}.
\ee
This gives rise to 
\be
p_{L,1}( \cos\theta_1- \beta ) &=& p_{L,2}( \cos\theta_2- \beta ) \label{re1},\\
p_{L,1}\sin\theta_1 &=& p_{L,2}\sin\theta_2. \label{re2}
\ee
One can solve Eqs. \ref{re1} and \ref{re2} for $\cos\theta_2$ and $p_{L,2}$, 
\be
\cos\theta_2 &=& -\frac{\beta^2 \cos\theta_1 -2 \beta+\cos\theta_1}{\beta^2-2 \beta \cos\theta_1+1}, \\
p_{L,2} &=& {p_0 \over \sqrt{1-\cos^2\theta_1 + \gamma^2 (\cos\theta_1 -\beta)^2} }\sqrt{1-\cos^2\theta_1 \over 1-\cos^2\theta_2}.
\ee
In summary, the final formula for the decay probability in the laboratory frame is as follows:
\be
P_L&=&\int d\beta f(\beta) dW {1 \over L_{L,1}} e^{-r_1/L_{L,1}} {1 \over L_{L,2}} e^{-r_2/L_{L,2}} dr_1 dr_2 \nonumber\\
  &=& \int_{-1}^1 d\beta f(\beta) \int_{-1}^1 {d\cos\theta_1 \over 2} \gamma (1-\beta \cos\theta_1) (\sin^2\theta_1 + \gamma^2 (\cos\theta_1 - \beta)^2)^{-3/2} \nonumber\\
&&\times(e^{-r_{1,min}/L_{L,1}}-e^{-r_{1,max}/L_{L,1}})(e^{-r_{2,min}/L_{L,2}}-e^{-r_{2,max}/L_{L,2}}),
\label{problab}
\ee
where
\be
L_{L,1} &=& p_{L,1} {\tau_{A'}\over m_{A'}} ={p_0 \over \sqrt{1-\cos^2\theta_1 + \gamma^2 (\cos\theta_1 -\beta)^2} }{\tau_{A'}\over m_{A'}},\\
L_{L,2} &=& p_{L,2} {\tau_{A'}\over m_{A'}} = {p_0 \over \sqrt{1-\cos^2\theta_1 + \gamma^2 (\cos\theta_1 -\beta)^2} }\sqrt{1-\cos^2\theta_1 \over 1-\cos^2\theta_2}\; {\tau_{A'}\over m_{A'}}.
\ee
We have used the delta function in the last step of Eq.~\ref{problab}. Note that $p_0\tau_{A'}/ m_{A'}$ is the decay length of 
the daughter particles in the rest frame of the parent particle. $r_{i,min}$ and $r_{i,max}$ are lower and upper boundaries of $r_i$ of the fiducial volume. $\theta_{1,min}$ and $\theta_{1,max}$ are lower and upper boundaries of $\theta_1$ of the fiducial region. $f(\beta)$ is the normalized velocity distribution  of the parent particle, {\em i.e.} $\int_{-1}^1 d\beta f(\beta) =1$.
We use $\texttt{MadGraph5\_aMC@NLO}$~\cite{Alwall:2014hca} to obtain $f(\beta)$ of the parent particle.

\bibliography{ref}

\providecommand{\href}[2]{#2}\begingroup\raggedright\begin{thebibliography}{10%
0}

\bibitem{ATLAS:2015xxx}
ATLAS collaboration, \emph{{Search for resonances decaying to photon pairs in
  3.2 fb$^{-1}$ of $pp$ collisions at $\sqrt{s}$ = 13 TeV with the ATLAS
  detector}}, \href{http://inspirehep.net/record/1410174/files/ATLAS-CONF-2015-081.pdf?version=1}{{\tt
  ATLAS-CONF-2015-081}}.

\bibitem{CMS:2015dxe}
{\scshape CMS} collaboration, \emph{{Search for new physics
  in high mass diphoton events in proton-proton collisions at 13TeV}}, \href{http://inspirehep.net/record/1409807/files/EXO-15-004-pas.pdf?version=1}{{\tt
  CMS-PAS-EXO-15-004}}.

\bibitem{Landau:1948kw}
L.~D. Landau, \emph{{On the angular momentum of a system of two photons}},
  \href{http://dx.doi.org/10.1016/B978-0-08-010586-4.50070-5}{\emph{Dokl. Akad.
  Nauk Ser. Fiz.} {\bf 60} (1948) 207--209}.

\bibitem{Yang:1950rg}
C.-N. Yang, \emph{{Selection Rules for the Dematerialization of a Particle Into
  Two Photons}},
  \href{http://dx.doi.org/10.1103/PhysRev.77.242}{\emph{Phys.Rev.} {\bf 77}
  (1950) 242--245}.

\bibitem{McDermott:2015sck}
S.~D. McDermott, P.~Meade and H.~Ramani, \emph{{Singlet Scalar Resonances and
  the Diphoton Excess}},  \href{http://arxiv.org/abs/1512.05326}{{\tt
  1512.05326}}.

\bibitem{Chao:2015nsm}
W.~Chao, \emph{{Symmetries Behind the 750 GeV Diphoton Excess}},
  \href{http://arxiv.org/abs/1512.06297}{{\tt 1512.06297}}.

\bibitem{Dev:2015vjd}
P.~S.~B. Dev, R.~N. Mohapatra and Y.~Zhang, \emph{{Quark Seesaw, Vectorlike
  Fermions and Diphoton Excess}},  \href{http://arxiv.org/abs/1512.08507}{{\tt
  1512.08507}}.

\bibitem{Murphy:2015kag}
C.~W. Murphy, \emph{{Vector Leptoquarks and the 750 GeV Diphoton Resonance at
  the LHC}},  \href{http://arxiv.org/abs/1512.06976}{{\tt 1512.06976}}.

\bibitem{Ellis:2015oso}
J.~Ellis, S.~A.~R. Ellis, J.~Quevillon, V.~Sanz and T.~You, \emph{{On the
  Interpretation of a Possible $\sim 750$ GeV Particle Decaying into $\gamma
  \gamma$}},  \href{http://arxiv.org/abs/1512.05327}{{\tt 1512.05327}}.

\bibitem{Wang:2015omi}
F.~Wang, W.~Wang, L.~Wu, J.~M. Yang and M.~Zhang, \emph{{Interpreting 750 GeV
  Diphoton Resonance in the NMSSM with Vector-like Particles}},
  \href{http://arxiv.org/abs/1512.08434}{{\tt 1512.08434}}.

\bibitem{Kawamura:2016idj}
J.~Kawamura and Y.~Omura, \emph{{Diphoton excess at 750 GeV and LHC constraints
  in models with vector-like particles}},
  \href{http://arxiv.org/abs/1601.07396}{{\tt 1601.07396}}.

\bibitem{Dolan:2016eki}
M.~J. Dolan, J.~L. Hewett, M.~Kr{\"a}mer and T.~G. Rizzo, \emph{{Simplified
  Models for Higgs Physics: Singlet Scalar and Vector-like Quark
  Phenomenology}},  \href{http://arxiv.org/abs/1601.07208}{{\tt 1601.07208}}.

\bibitem{Dutta:2016jqn}
B.~Dutta, Y.~Gao, T.~Ghosh, I.~Gogoladze, T.~Li, Q.~Shafi et~al.,
  \emph{{Diphoton Excess in Consistent Supersymmetric SU(5) Models with
  Vector-like Particles}},  \href{http://arxiv.org/abs/1601.00866}{{\tt
  1601.00866}}.

\bibitem{Ge:2016xcq}
S.-F. Ge, H.-J. He, J.~Ren and Z.-Z. Xianyu, \emph{{Realizing Dark Matter and
  Higgs Inflation in Light of LHC Diphoton Excess}},
  \href{http://arxiv.org/abs/1602.01801}{{\tt 1602.01801}}.

\bibitem{Alves:2015jgx}
A.~Alves, A.~G. Dias and K.~Sinha, \emph{{The 750 GeV $S$-cion: Where else
  should we look for it?}},  \href{http://arxiv.org/abs/1512.06091}{{\tt
  1512.06091}}.

\bibitem{Buttazzo:2015txu}
D.~Buttazzo, A.~Greljo and D.~Marzocca, \emph{{Knocking on New Physics' door
  with a Scalar Resonance}},  \href{http://arxiv.org/abs/1512.04929}{{\tt
  1512.04929}}.

\bibitem{Bai:2015nbs}
Y.~Bai, J.~Berger and R.~Lu, \emph{{A 750 GeV Dark Pion: Cousin of a Dark
  G-parity-odd WIMP}},  \href{http://arxiv.org/abs/1512.05779}{{\tt
  1512.05779}}.

\bibitem{Ding:2015rxx}
R.~Ding, L.~Huang, T.~Li and B.~Zhu, \emph{{Interpreting $750$ GeV Diphoton
  Excess with R-parity Violation Supersymmetry}},
  \href{http://arxiv.org/abs/1512.06560}{{\tt 1512.06560}}.

\bibitem{Huang:2015evq}
F.~P. Huang, C.~S. Li, Z.~L. Liu and Y.~Wang, \emph{{750 GeV Diphoton Excess
  from Cascade Decay}},  \href{http://arxiv.org/abs/1512.06732}{{\tt
  1512.06732}}.

\bibitem{Berthier:2015vbb}
L.~Berthier, J.~M. Cline, W.~Shepherd and M.~Trott, \emph{{Effective
  interpretations of a diphoton excess}},
  \href{http://arxiv.org/abs/1512.06799}{{\tt 1512.06799}}.

\bibitem{Gu:2015lxj}
J.~Gu and Z.~Liu, \emph{{Running after Diphoton}},
  \href{http://arxiv.org/abs/1512.07624}{{\tt 1512.07624}}.

\bibitem{Cai:2015hzc}
C.~Cai, Z.-H. Yu and H.-H. Zhang, \emph{{The 750 GeV diphoton resonance as a
  singlet scalar in an extra dimensional model}},
  \href{http://arxiv.org/abs/1512.08440}{{\tt 1512.08440}}.

\bibitem{Son:2015vfl}
M.~Son and A.~Urbano, \emph{{A new scalar resonance at 750 GeV: Towards a proof
  of concept in favor of strongly interacting theories}},
  \href{http://arxiv.org/abs/1512.08307}{{\tt 1512.08307}}.

\bibitem{Cheung:2015cug}
K.~Cheung, P.~Ko, J.~S. Lee, J.~Park and P.-Y. Tseng, \emph{{A Higgcision study
  on the 750 GeV Di-photon Resonance and 125 GeV SM Higgs boson with the
  Higgs-Singlet Mixing}},  \href{http://arxiv.org/abs/1512.07853}{{\tt
  1512.07853}}.

\bibitem{Badziak:2015zez}
M.~Badziak, \emph{{Interpreting the 750 GeV diphoton excess in minimal
  extensions of Two-Higgs-Doublet models}},
  \href{http://arxiv.org/abs/1512.07497}{{\tt 1512.07497}}.

\bibitem{Han:2015qqj}
X.-F. Han and L.~Wang, \emph{{Implication of the 750 GeV diphoton resonance on
  two-Higgs-doublet model and its extensions with Higgs field}},
  \href{http://dx.doi.org/10.1103/PhysRevD.93.055027}{\emph{Phys. Rev.} {\bf
  D93} (2016) 055027}, [\href{http://arxiv.org/abs/1512.06587}{{\tt
  1512.06587}}].

\bibitem{Moretti:2015pbj}
S.~Moretti and K.~Yagyu, \emph{{The 750 GeV diphoton excess and its explanation
  in 2-Higgs Doublet Models with a real inert scalar multiplet}},
  \href{http://arxiv.org/abs/1512.07462}{{\tt 1512.07462}}.

\bibitem{Huang:2015rkj}
W.-C. Huang, Y.-L.~S. Tsai and T.-C. Yuan, \emph{{Gauged Two Higgs Doublet
  Model confronts the LHC 750 GeV di-photon anomaly}},
  \href{http://arxiv.org/abs/1512.07268}{{\tt 1512.07268}}.

\bibitem{Chakraborty:2015jvs}
I.~Chakraborty and A.~Kundu, \emph{{Diphoton excess at 750 GeV: Singlet scalars
  confront triviality}},  \href{http://arxiv.org/abs/1512.06508}{{\tt
  1512.06508}}.

\bibitem{Han:2015dlp}
H.~Han, S.~Wang and S.~Zheng, \emph{{Scalar Explanation of Diphoton Excess at
  LHC}},  \href{http://arxiv.org/abs/1512.06562}{{\tt 1512.06562}}.

\bibitem{Ghosh:2015apa}
S.~Ghosh, A.~Kundu and S.~Ray, \emph{{On the potential of a singlet scalar
  enhanced Standard Model}},  \href{http://arxiv.org/abs/1512.05786}{{\tt
  1512.05786}}.

\bibitem{Kanemura:2015vcb}
S.~Kanemura, N.~Machida, S.~Odori and T.~Shindou, \emph{{Diphoton excess at 750
  GeV in an extended scalar sector}},
  \href{http://arxiv.org/abs/1512.09053}{{\tt 1512.09053}}.

\bibitem{Chiang:2015tqz}
C.-W. Chiang, M.~Ibe and T.~T. Yanagida, \emph{{Revisiting Scalar Quark Hidden
  Sector in Light of 750-GeV Diphoton Resonance}},
  \href{http://arxiv.org/abs/1512.08895}{{\tt 1512.08895}}.

\bibitem{Kobakhidze:2015ldh}
A.~Kobakhidze, F.~Wang, L.~Wu, J.~M. Yang and M.~Zhang, \emph{{LHC 750 GeV
  diphoton resonance explained as a heavy scalar in top-seesaw model}},
  \href{http://arxiv.org/abs/1512.05585}{{\tt 1512.05585}}.

\bibitem{Altmannshofer:2015xfo}
W.~Altmannshofer, J.~Galloway, S.~Gori, A.~L. Kagan, A.~Martin and J.~Zupan,
  \emph{{On the 750 GeV di-photon excess}},
  \href{http://arxiv.org/abs/1512.07616}{{\tt 1512.07616}}.

\bibitem{Ahmed:2015uqt}
A.~Ahmed, B.~M. Dillon, B.~Grzadkowski, J.~F. Gunion and Y.~Jiang,
  \emph{{Higgs-radion interpretation of 750 GeV di-photon excess at the LHC}},
  \href{http://arxiv.org/abs/1512.05771}{{\tt 1512.05771}}.

\bibitem{Falkowski:2015swt}
A.~Falkowski, O.~Slone and T.~Volansky, \emph{{Phenomenology of a 750 GeV
  Singlet}}, \href{http://dx.doi.org/10.1007/JHEP02(2016)152}{\emph{JHEP} {\bf
  02} (2016) 152}, [\href{http://arxiv.org/abs/1512.05777}{{\tt 1512.05777}}].

\bibitem{Agrawal:2015dbf}
P.~Agrawal, J.~Fan, B.~Heidenreich, M.~Reece and M.~Strassler,
  \emph{{Experimental Considerations Motivated by the Diphoton Excess at the
  LHC}},  \href{http://arxiv.org/abs/1512.05775}{{\tt 1512.05775}}.

\bibitem{Dasgupta:2016wxw}
B.~Dasgupta, J.~Kopp and P.~Schwaller, \emph{{Photons, Photon Jets and Dark
  Photons at 750\,GeV and Beyond}},
  \href{http://arxiv.org/abs/1602.04692}{{\tt 1602.04692}}.

\bibitem{Draper:2012xt}
P.~Draper and D.~McKeen, \emph{{Diphotons from Tetraphotons in the Decay of a
  125 GeV Higgs at the LHC}},
  \href{http://dx.doi.org/10.1103/PhysRevD.85.115023}{\emph{Phys.Rev.} {\bf
  D85} (2012) 115023}, [\href{http://arxiv.org/abs/1204.1061}{{\tt
  1204.1061}}].

\bibitem{Bi:2015lcf}
X.-J. Bi, R.~Ding, Y.~Fan, L.~Huang, C.~Li, T.~Li et~al., \emph{{A Promising
  Interpretation of Diphoton Resonance at 750 GeV}},
  \href{http://arxiv.org/abs/1512.08497}{{\tt 1512.08497}}.

\bibitem{Boehm:2002yz}
C.~Boehm, T.~A. Ensslin and J.~Silk, \emph{{Can Annihilating dark matter be
  lighter than a few GeVs?}},
  \href{http://dx.doi.org/10.1088/0954-3899/30/3/004}{\emph{J. Phys.} {\bf G30}
  (2004) 279--286}, [\href{http://arxiv.org/abs/astro-ph/0208458}{{\tt
  astro-ph/0208458}}].

\bibitem{Boehm:2003hm}
C.~Boehm and P.~Fayet, \emph{{Scalar dark matter candidates}},
  \href{http://dx.doi.org/10.1016/j.nuclphysb.2004.01.015}{\emph{Nucl. Phys.}
  {\bf B683} (2004) 219--263}, [\href{http://arxiv.org/abs/hep-ph/0305261}{{\tt
  hep-ph/0305261}}].

\bibitem{Pospelov:2007mp}
M.~Pospelov, A.~Ritz and M.~B. Voloshin, \emph{{Secluded WIMP Dark Matter}},
  \href{http://dx.doi.org/10.1016/j.physletb.2008.02.052}{\emph{Phys. Lett.}
  {\bf B662} (2008) 53--61}, [\href{http://arxiv.org/abs/0711.4866}{{\tt
  0711.4866}}].

\bibitem{ArkaniHamed:2008qn}
N.~Arkani-Hamed, D.~P. Finkbeiner, T.~R. Slatyer and N.~Weiner, \emph{{A Theory
  of Dark Matter}},
  \href{http://dx.doi.org/10.1103/PhysRevD.79.015014}{\emph{Phys. Rev.} {\bf
  D79} (2009) 015014}, [\href{http://arxiv.org/abs/0810.0713}{{\tt
  0810.0713}}].

\bibitem{Essig:2013lka}
R.~Essig et~al., \emph{{Working Group Report: New Light Weakly Coupled
  Particles}},  in \emph{{Community Summer Study 2013: Snowmass on the
  Mississippi (CSS2013) Minneapolis, MN, USA, July 29-August 6, 2013}}, 2013.
\newblock \href{http://arxiv.org/abs/1311.0029}{{\tt 1311.0029}}.

\bibitem{Hewett:2012ns}
J.~Hewett, H.~Weerts, R.~Brock, J.~Butler, B.~Casey et~al., \emph{{Fundamental
  Physics at the Intensity Frontier}},
  \href{http://arxiv.org/abs/1205.2671}{{\tt 1205.2671}}.

\bibitem{Strassler:2006im}
M.~J. Strassler and K.~M. Zurek, \emph{{Echoes of a hidden valley at hadron
  colliders}},
  \href{http://dx.doi.org/10.1016/j.physletb.2007.06.055}{\emph{Phys.Lett.}
  {\bf B651} (2007) 374--379}, [\href{http://arxiv.org/abs/hep-ph/0604261}{{\tt
  hep-ph/0604261}}].

\bibitem{Chang:2015sdy}
J.~Chang, K.~Cheung and C.-T. Lu, \emph{{Interpreting the 750 GeV Di-photon
  Resonance using photon-jets in Hidden-Valley-like models}},
  \href{http://arxiv.org/abs/1512.06671}{{\tt 1512.06671}}.

\bibitem{Chang:2015bzc}
S.~Chang, \emph{{A Simple $U(1)$ Gauge Theory Explanation of the Diphoton
  Excess}},  \href{http://arxiv.org/abs/1512.06426}{{\tt 1512.06426}}.

\bibitem{Bian:2015kjt}
L.~Bian, N.~Chen, D.~Liu and J.~Shu, \emph{{A hidden confining world on the 750
  GeV diphoton excess}},  \href{http://arxiv.org/abs/1512.05759}{{\tt
  1512.05759}}.

\bibitem{Knapen:2015dap}
S.~Knapen, T.~Melia, M.~Papucci and K.~Zurek, \emph{{Rays of light from the
  LHC}},  \href{http://arxiv.org/abs/1512.04928}{{\tt 1512.04928}}.

\bibitem{Dey:2015bur}
U.~K. Dey, S.~Mohanty and G.~Tomar, \emph{{750 GeV resonance in the Dark
  Left-Right Model}},  \href{http://arxiv.org/abs/1512.07212}{{\tt
  1512.07212}}.

\bibitem{Davoudiasl:2015cuo}
H.~Davoudiasl and C.~Zhang, \emph{{A 750 GeV Messenger of Dark Conformal
  Symmetry Breaking}},  \href{http://arxiv.org/abs/1512.07672}{{\tt
  1512.07672}}.

\bibitem{Das:2015enc}
K.~Das and S.~K. Rai, \emph{{The 750 GeV Diphoton excess in a $U(1)$ hidden
  symmetry model}},  \href{http://arxiv.org/abs/1512.07789}{{\tt 1512.07789}}.

\bibitem{Kaneta:2015qpf}
K.~Kaneta, S.~Kang and H.-S. Lee, \emph{{Diphoton excess at the LHC Run 2 and
  its implications for a new heavy gauge boson}},
  \href{http://arxiv.org/abs/1512.09129}{{\tt 1512.09129}}.

\bibitem{Jiang:2015oms}
Y.~Jiang, Y.-Y. Li and T.~Liu, \emph{{750 GeV Resonance in the Gauged
  $U(1)'$-Extended MSSM}},  \href{http://arxiv.org/abs/1512.09127}{{\tt
  1512.09127}}.

\bibitem{Yu:2016lof}
J.-H. Yu, \emph{{Hidden Gauged U(1) Model: Unifying Scotogenic Neutrino and
  Flavor Dark Matter}},  \href{http://arxiv.org/abs/1601.02609}{{\tt
  1601.02609}}.

\bibitem{An:2015cgp}
H.~An, C.~Cheung and Y.~Zhang, \emph{{Broad Diphotons from Narrow States}},
  \href{http://arxiv.org/abs/1512.08378}{{\tt 1512.08378}}.

\bibitem{Bjorken:2009mm}
J.~D. Bjorken, R.~Essig, P.~Schuster and N.~Toro, \emph{{New Fixed-Target
  Experiments to Search for Dark Gauge Forces}},
  \href{http://dx.doi.org/10.1103/PhysRevD.80.075018}{\emph{Phys. Rev.} {\bf
  D80} (2009) 075018}.

\bibitem{Bjorken:1988as}
J.~D. Bjorken et~al., \emph{{Search for Neutral Metastable Penetrating
  Particles Produced in the SLAC Beam Dump}},
  \href{http://dx.doi.org/10.1103/PhysRevD.38.3375}{\emph{Phys. Rev.} {\bf D38}
  (1988) 3375}.

\bibitem{Riordan:1987aw}
E.~M. Riordan et~al., \emph{{A Search for Short Lived Axions in an Electron
  Beam Dump Experiment}},
  \href{http://dx.doi.org/10.1103/PhysRevLett.59.755}{\emph{Phys. Rev. Lett.}
  {\bf 59} (1987) 755}.

\bibitem{Bross:1989mp}
A.~Bross et~al., \emph{{A Search for Shortlived Particles Produced in an
  Electron Beam Dump}},
  \href{http://dx.doi.org/10.1103/PhysRevLett.67.2942}{\emph{Phys. Rev. Lett.}
  {\bf 67} (1991) 2942--2945}.

\bibitem{Batell:2009yf}
B.~Batell, M.~Pospelov and A.~Ritz, \emph{{Probing a Secluded U(1) at
  B-factories}},
  \href{http://dx.doi.org/10.1103/PhysRevD.79.115008}{\emph{Phys. Rev.} {\bf
  D79} (2009) 115008}.

\bibitem{Strassler:2006qa}
M.~J. Strassler, \emph{{Possible effects of a hidden valley on supersymmetric
  phenomenology}},  \href{http://arxiv.org/abs/hep-ph/0607160}{{\tt
  hep-ph/0607160}}.

\bibitem{Essig:2009nc}
R.~Essig, P.~Schuster and N.~Toro, \emph{{Probing Dark Forces and Light Hidden
  Sectors at Low-Energy e+e- Colliders}},
  \href{http://dx.doi.org/10.1103/PhysRevD.80.015003}{\emph{Phys. Rev.} {\bf
  D80} (2009) 015003}, [\href{http://arxiv.org/abs/0903.3941}{{\tt
  0903.3941}}].

\bibitem{Freytsis:2009bh}
M.~Freytsis, G.~Ovanesyan and J.~Thaler, \emph{{Dark Force Detection in Low
  Energy e-p Collisions}},
  \href{http://dx.doi.org/10.1007/JHEP01(2010)111}{\emph{JHEP} {\bf 1001}
  (2010) 111}, [\href{http://arxiv.org/abs/0909.2862}{{\tt 0909.2862}}].

\bibitem{Essig:2010xa}
R.~Essig, P.~Schuster, N.~Toro and B.~Wojtsekhowski, \emph{{An Electron Fixed
  Target Experiment to Search for a New Vector Boson A' Decaying to e+e-}},
  \href{http://dx.doi.org/10.1007/JHEP02(2011)009}{\emph{JHEP} {\bf 1102}
  (2011) 009}, [\href{http://arxiv.org/abs/1001.2557}{{\tt 1001.2557}}].

\bibitem{Blumlein:2011mv}
J.~Blumlein and J.~Brunner, \emph{{New Exclusion Limits for Dark Gauge Forces
  from Beam-Dump Data}},
  \href{http://dx.doi.org/10.1016/j.physletb.2011.05.046}{\emph{Phys.Lett.}
  {\bf B701} (2011) 155--159}, [\href{http://arxiv.org/abs/1104.2747}{{\tt
  1104.2747}}].

\bibitem{Andreas:2012mt}
S.~Andreas, C.~Niebuhr and A.~Ringwald, \emph{{New Limits on Hidden Photons
  from Past Electron Beam Dumps}},
  \href{http://dx.doi.org/10.1103/PhysRevD.86.095019}{\emph{Phys.Rev.} {\bf
  D86} (2012) 095019}, [\href{http://arxiv.org/abs/1209.6083}{{\tt
  1209.6083}}].

\bibitem{Pospelov:2008zw}
M.~Pospelov, \emph{{Secluded U(1) below the weak scale}},
  \href{http://dx.doi.org/10.1103/PhysRevD.80.095002}{\emph{Phys. Rev.} {\bf
  D80} (2009) 095002}, [\href{http://arxiv.org/abs/0811.1030}{{\tt
  0811.1030}}].

\bibitem{Reece:2009un}
M.~Reece and L.-T. Wang, \emph{{Searching for the light dark gauge boson in
  GeV-scale experiments}},
  \href{http://dx.doi.org/10.1088/1126-6708/2009/07/051}{\emph{JHEP} {\bf 07}
  (2009) 051}.

\bibitem{Aubert:2009cp}
{\scshape BaBar Collaboration} collaboration, B.~Aubert et~al., \emph{{Search
  for Dimuon Decays of a Light Scalar Boson in Radiative Transitions $\Upsilon
  \rightarrow \gamma A^0$}},
  \href{http://dx.doi.org/10.1103/PhysRevLett.103.081803}{\emph{Phys.Rev.Lett.}
  {\bf 103} (2009) 081803}, [\href{http://arxiv.org/abs/0905.4539}{{\tt
  0905.4539}}].

\bibitem{Hook:2010tw}
A.~Hook, E.~Izaguirre and J.~G. Wacker, \emph{{Model Independent Bounds on
  Kinetic Mixing}}, \href{http://dx.doi.org/10.1155/2011/859762}{\emph{Adv.High
  Energy Phys.} {\bf 2011} (2011) 859762},
  [\href{http://arxiv.org/abs/1006.0973}{{\tt 1006.0973}}].

\bibitem{Babusci:2012cr}
{\scshape KLOE-2 Collaboration} collaboration, D.~Babusci et~al., \emph{{Limit
  on the production of a light vector gauge boson in phi meson decays with the
  KLOE detector}},
  \href{http://dx.doi.org/10.1016/j.physletb.2013.01.067}{\emph{Phys.Lett.}
  {\bf B720} (2013) 111--115}, [\href{http://arxiv.org/abs/1210.3927}{{\tt
  1210.3927}}].

\bibitem{Archilli:2011zc}
F.~Archilli, D.~Babusci, D.~Badoni, I.~Balwierz, G.~Bencivenni et~al.,
  \emph{{Search for a vector gauge boson in phi meson decays with the KLOE
  detector}},
  \href{http://dx.doi.org/10.1016/j.physletb.2011.11.033}{\emph{Phys.Lett.}
  {\bf B706} (2012) 251--255}, [\href{http://arxiv.org/abs/1110.0411}{{\tt
  1110.0411}}].

\bibitem{Abrahamyan:2011gv}
{\scshape APEX} collaboration, S.~Abrahamyan et~al., \emph{{Search for a new
  gauge boson in the $A'$ Experiment (APEX)}},
  \href{http://dx.doi.org/10.1103/PhysRevLett.107.191804}{\emph{Phys. Rev.
  Lett.} {\bf 107} (2011) 191804}, [\href{http://arxiv.org/abs/1108.2750}{{\tt
  1108.2750}}].

\bibitem{Merkel:2014avp}
H.~Merkel, P.~Achenbach, C.~A. Gayoso, T.~Beranek, J.~Bericic et~al.,
  \emph{{Search for light massive gauge bosons as an explanation of the
  $(g-2)_\mu$ anomaly at MAMI}},  \href{http://arxiv.org/abs/1404.5502}{{\tt
  1404.5502}}.

\bibitem{Dent:2012mx}
J.~B. Dent, F.~Ferrer and L.~M. Krauss, \emph{{Constraints on Light Hidden
  Sector Gauge Bosons from Supernova Cooling}},
  \href{http://arxiv.org/abs/1201.2683}{{\tt 1201.2683}}.

\bibitem{Davoudiasl:2012ig}
H.~Davoudiasl, H.-S. Lee and W.~J. Marciano, \emph{{Dark Side of Higgs Diphoton
  Decays and Muon g-2}},
  \href{http://dx.doi.org/10.1103/PhysRevD.86.095009}{\emph{Phys.Rev.} {\bf
  D86} (2012) 095009}, [\href{http://arxiv.org/abs/1208.2973}{{\tt
  1208.2973}}].

\bibitem{Davoudiasl:2012ag}
H.~Davoudiasl, H.-S. Lee and W.~J. Marciano, \emph{{'Dark' Z implications for
  Parity Violation, Rare Meson Decays, and Higgs Physics}},
  \href{http://dx.doi.org/10.1103/PhysRevD.85.115019}{\emph{Phys.Rev.} {\bf
  D85} (2012) 115019}, [\href{http://arxiv.org/abs/1203.2947}{{\tt
  1203.2947}}].

\bibitem{Davoudiasl:2013aya}
H.~Davoudiasl, H.-S. Lee, I.~Lewis and W.~J. Marciano, \emph{{Higgs Decays as a
  Window into the Dark Sector}},  \href{http://arxiv.org/abs/1304.4935}{{\tt
  1304.4935}}.

\bibitem{Endo:2012hp}
M.~Endo, K.~Hamaguchi and G.~Mishima, \emph{{Constraints on Hidden Photon
  Models from Electron g-2 and Hydrogen Spectroscopy}},
  \href{http://dx.doi.org/10.1103/PhysRevD.86.095029}{\emph{Phys.Rev.} {\bf
  D86} (2012) 095029}, [\href{http://arxiv.org/abs/1209.2558}{{\tt
  1209.2558}}].

\bibitem{Balewski:2013oza}
J.~Balewski, J.~Bernauer, W.~Bertozzi, J.~Bessuille, B.~Buck et~al.,
  \emph{{DarkLight: A Search for Dark Forces at the Jefferson Laboratory
  Free-Electron Laser Facility}},  \href{http://arxiv.org/abs/1307.4432}{{\tt
  1307.4432}}.

\bibitem{Adlarson:2013eza}
{\scshape WASA-at-COSY} collaboration, P.~Adlarson et~al., \emph{{Search for a
  dark photon in the $\pi^0 \to e^+e^-\gamma$ decay}},
  \href{http://dx.doi.org/10.1016/j.physletb.2013.08.055}{\emph{Phys.Lett.}
  {\bf B726} (2013) 187--193}, [\href{http://arxiv.org/abs/1304.0671}{{\tt
  1304.0671}}].

\bibitem{Agakishiev:2013fwl}
{\scshape HADES} collaboration, G.~Agakishiev et~al., \emph{{Searching a Dark
  Photon with HADES}},
  \href{http://dx.doi.org/10.1016/j.physletb.2014.02.035}{\emph{Phys.Lett.}
  {\bf B731} (2014) 265--271}, [\href{http://arxiv.org/abs/1311.0216}{{\tt
  1311.0216}}].

\bibitem{Andreas:2013lya}
S.~Andreas, S.~Donskov, P.~Crivelli, A.~Gardikiotis, S.~Gninenko et~al.,
  \emph{{Proposal for an Experiment to Search for Light Dark Matter at the
  SPS}},  \href{http://arxiv.org/abs/1312.3309}{{\tt 1312.3309}}.

\bibitem{Battaglieri:2014hga}
M.~Battaglieri, S.~Boyarinov, S.~Bueltmann, V.~Burkert, A.~Celentano et~al.,
  \emph{{The Heavy Photon Search Test Detector}},
  \href{http://arxiv.org/abs/1406.6115}{{\tt 1406.6115}}.

\bibitem{Lees:2014xha}
{\scshape BaBar} collaboration, J.~Lees et~al., \emph{{Search for a dark photon
  in e+e- collisions at BABAR}},  \href{http://arxiv.org/abs/1406.2980}{{\tt
  1406.2980}}.

\bibitem{Adare:2014ega}
A.~Adare, S.~Afanasiev, C.~Aidala, N.~Ajitanand, Y.~Akiba et~al.,
  \emph{{Closing the Door for Dark Photons as the Explanation for the Muon g-2
  Anomaly}},  \href{http://arxiv.org/abs/1409.0851}{{\tt 1409.0851}}.

\bibitem{Kazanas:2014mca}
D.~Kazanas, R.~N. Mohapatra, S.~Nussinov, V.~Teplitz and Y.~Zhang,
  \emph{{Supernova Bounds on the Dark Photon Using its Electromagnetic Decay}},
   \href{http://arxiv.org/abs/1410.0221}{{\tt 1410.0221}}.

\bibitem{Blumlein:2013cua}
J.~Bl{\"u}mlein and J.~Brunner, \emph{{New Exclusion Limits on Dark Gauge
  Forces from Proton Bremsstrahlung in Beam-Dump Data}},
  \href{http://dx.doi.org/10.1016/j.physletb.2014.02.029}{\emph{Phys.Lett.}
  {\bf B731} (2014) 320--326}, [\href{http://arxiv.org/abs/1311.3870}{{\tt
  1311.3870}}].

\bibitem{CERNNA48/2:2015lha}
{\scshape CERN NA48/2} collaboration, C.~N. Collaboration, \emph{{Search for
  the dark photon in $\pi^0$ decays}},
  \href{http://arxiv.org/abs/1504.00607}{{\tt 1504.00607}}.

\bibitem{Stepanyan:2013tma}
{\scshape HPS} collaboration, S.~Stepanyan, \emph{{Heavy photon search
  experiment at JLAB}}, \href{http://dx.doi.org/10.1063/1.4829398}{\emph{AIP
  Conf.Proc.} {\bf 1563} (2013) 155--158}.

\bibitem{Echenard:2014lma}
B.~Echenard, R.~Essig and Y.-M. Zhong, \emph{{Projections for Dark Photon
  Searches at Mu3e}},
  \href{http://dx.doi.org/10.1007/JHEP01(2015)113}{\emph{JHEP} {\bf 01} (2015)
  113}, [\href{http://arxiv.org/abs/1411.1770}{{\tt 1411.1770}}].

\bibitem{Alekhin:2015byh}
S.~Alekhin et~al., \emph{{A facility to Search for Hidden Particles at the CERN
  SPS: the SHiP physics case}},  \href{http://arxiv.org/abs/1504.04855}{{\tt
  1504.04855}}.

\bibitem{Ilten:2015hya}
P.~Ilten, J.~Thaler, M.~Williams and W.~Xue, \emph{{Dark photons from charm
  mesons at LHCb}},
  \href{http://dx.doi.org/10.1103/PhysRevD.92.115017}{\emph{Phys. Rev.} {\bf
  D92} (2015) 115017}, [\href{http://arxiv.org/abs/1509.06765}{{\tt
  1509.06765}}].

\bibitem{Holdom:1985ag}
B.~Holdom, \emph{{Two U(1)'s and Epsilon Charge Shifts}},
  \href{http://dx.doi.org/10.1016/0370-2693(86)91377-8}{\emph{Phys.Lett.} {\bf
  B166} (1986) 196}.

\bibitem{Galison:1983pa}
P.~Galison and A.~Manohar, \emph{{Two Z's or Not Two Z's?}},
  \href{http://dx.doi.org/10.1016/0370-2693(84)91161-4}{\emph{Phys.Lett.} {\bf
  B136} (1984) 279}.

\bibitem{Boehm:2003ha}
C.~Boehm, P.~Fayet and J.~Silk, \emph{{Light and heavy dark matter particles}},
  \href{http://dx.doi.org/10.1103/PhysRevD.69.101302}{\emph{Phys. Rev.} {\bf
  D69} (2004) 101302}, [\href{http://arxiv.org/abs/hep-ph/0311143}{{\tt
  hep-ph/0311143}}].

\bibitem{Gunion:1989we}
J.~F. Gunion, H.~E. Haber, G.~L. Kane and S.~Dawson, \emph{{The Higgs Hunter's
  Guide}}, {\emph{Front.Phys.} {\bf 80} (2000) 1--448}.

\bibitem{Pospelov:2008jd}
M.~Pospelov and A.~Ritz, \emph{{Astrophysical Signatures of Secluded Dark
  Matter}},
  \href{http://dx.doi.org/10.1016/j.physletb.2008.12.012}{\emph{Phys.Lett.}
  {\bf B671} (2009) 391--397}, [\href{http://arxiv.org/abs/0810.1502}{{\tt
  0810.1502}}].

\bibitem{Toro:2012sv}
N.~Toro and I.~Yavin, \emph{{Multiphotons and photon jets from new heavy vector
  bosons}}, \href{http://dx.doi.org/10.1103/PhysRevD.86.055005}{\emph{Phys.
  Rev.} {\bf D86} (2012) 055005}, [\href{http://arxiv.org/abs/1202.6377}{{\tt
  1202.6377}}].

\bibitem{Chala:2015cev}
M.~Chala, M.~Duerr, F.~Kahlhoefer and K.~Schmidt-Hoberg, \emph{{Tricking
  Landau–Yang: How to obtain the diphoton excess from a vector resonance}},
  \href{http://dx.doi.org/10.1016/j.physletb.2016.02.006}{\emph{Phys. Lett.}
  {\bf B755} (2016) 145--149}, [\href{http://arxiv.org/abs/1512.06833}{{\tt
  1512.06833}}].

\bibitem{ATLAS:2012ana}
{\scshape ATLAS} collaboration, \emph{{Measurements of the photon
  identification efficiency with the ATLAS detector using 4.9 fb$^{-1}$ of pp
  collision data collected in 2011}},\href{http://inspirehep.net/record/1204294?ln=en}{{\tt
  ATLAS-CONF-2012-123}}.

\bibitem{Rohne201318}
O.~R{\o}hne, \emph{Overview of the \{ATLAS\} insertable b-layer (ibl) project},
  \href{http://dx.doi.org/http://dx.doi.org/10.1016/j.nima.2013.05.032}{\emph{%
Nuclear Instruments and Methods in Physics Research Section A: Accelerators,
  Spectrometers, Detectors and Associated Equipment} {\bf 731} (2013) 18 --
  24}.

\bibitem{Aad:2009wy}
{\scshape ATLAS} collaboration, G.~Aad et~al., \emph{{Expected Performance of
  the ATLAS Experiment - Detector, Trigger and Physics}},
  \href{http://arxiv.org/abs/0901.0512}{{\tt 0901.0512}}.

\bibitem{Aad:2014tda}
{\scshape ATLAS} collaboration, G.~Aad et~al., \emph{{Search for new phenomena
  in events with a photon and missing transverse momentum in $pp$ collisions at
  $\sqrt{s}=8$ TeV with the ATLAS detector}},
  \href{http://dx.doi.org/10.1103/PhysRevD.92.059903,
  10.1103/PhysRevD.91.012008}{\emph{Phys. Rev.} {\bf D91} (2015) 012008},
  [\href{http://arxiv.org/abs/1411.1559}{{\tt 1411.1559}}].

\bibitem{Balewski:2014pxa}
J.~Balewski, J.~Bernauer, J.~Bessuille, R.~Corliss, R.~Cowan et~al., \emph{{The
  DarkLight Experiment: A Precision Search for New Physics at Low Energies}},
  \href{http://arxiv.org/abs/1412.4717}{{\tt 1412.4717}}.

\bibitem{Abe:2010gxa}
{\scshape Belle-II} collaboration, T.~Abe et~al., \emph{{Belle II Technical
  Design Report}},  \href{http://arxiv.org/abs/1011.0352}{{\tt 1011.0352}}.

\bibitem{Bi:2016gca}
X.-J. Bi, Z.~Kang, P.~Ko, J.~Li and T.~Li, \emph{{ADMonium: Asymmetric Dark
  Matter Bound State}},  \href{http://arxiv.org/abs/1602.08816}{{\tt
  1602.08816}}.

\bibitem{wang}
Y.~Tsai, L.-T. Wang and Y.~Zhao, \emph{{Faking The Diphoton Excess by Displaced
  Dark Photon Decays}},  \href{http://arxiv.org/abs/1603.00024}{{\tt
  1603.00024}}.

\bibitem{Alwall:2014hca}
J.~Alwall, R.~Frederix, S.~Frixione, V.~Hirschi, F.~Maltoni et~al., \emph{{The
  automated computation of tree-level and next-to-leading order differential
  cross sections, and their matching to parton shower simulations}},
  \href{http://dx.doi.org/10.1007/JHEP07(2014)079}{\emph{JHEP} {\bf 1407}
  (2014) 079}, [\href{http://arxiv.org/abs/1405.0301}{{\tt 1405.0301}}].

\end{thebibliography}\endgroup
\bibliographystyle{JHEP}

%\end{thebibliography}
\end{document}